\documentclass[review,12pt]{elsarticle}

\usepackage{amssymb}
\usepackage{amsmath}
\usepackage{amsfonts}
\usepackage{algorithmic}
\usepackage{graphicx}
\usepackage{textcomp}
\usepackage{xcolor}
\usepackage{hyperref}
\usepackage{mathrsfs}
\usepackage{bm}
\usepackage{multirow}
\usepackage{txfonts}
\usepackage{slashbox}
\usepackage{subcaption}
\usepackage{makecell}
\usepackage{lscape}

\newtheorem{theorem}{Theorem}
\newtheorem{lemma}{Lemma}

\def\beq{\begin{equation}}
\def\eeq{\end{equation}}
\def\bea{\begin{eqnarray}}
\def\eea{\end{eqnarray}}
\def\ba{\begin{array}}
\def\ea{\end{array}}

\def\bitem{\begin{itemize}}
\def\eitem{\end{itemize}}
\def\ben{\begin{enumerate}}
\def\een{\end{enumerate}}

\def\ie{{\it i.e.,\ \/}}

\definecolor{bgrd}{rgb}{1,1,1}
\definecolor{gray}{rgb}{0.5,0.5,0.5}
\definecolor{dkr}{rgb}{0.7,0.1,0.2}
\definecolor{dkb}{rgb}{0.1,0.1,0.8}

\renewcommand{\thefootnote}{\fnsymbol{footnote}}

\makeatletter
\newdimen{\captionwidth}
\long\def\@makecaption#1#2{%
\captionwidth .9\hsize
\vskip 10pt%
\setbox\@tempboxa\hbox{#1: #2}%
  \ifdim \wd\@tempboxa >\captionwidth%
    \setbox\@tempboxa\hbox{#1:\hspace*{.5em}}%
    \hfil\parbox{\captionwidth}{\raggedright\hangindent \wd\@tempboxa%
    \hangafter=1\unhbox\@tempboxa#2}\hfill%
  \else\centerline{\box\@tempboxa}%
  \fi
}
\makeatother

\newcommand{\mbbE}{\mathbb{E}}

\newcommand{\mbbR}{\mathbb{R}}

\def\nubf{\hbox{\boldmath$\nu$\unboldmath}}

\def\xbf{{\bf x}}

\def\xbf{{\bf x}}

\def\Ubf{{\bf U}}
\def\Vbf{{\bf V}}

\def\Xbf{{\bf X}}
\def\Ybf{{\bf Y}}

\def\Uc{{\cal U}}
\def\Vc{{\cal V}}

\def\xbf{{\bm x}}
\def\Xbf{{\bm X}}

\journal{International Journal of Forecasting}

\begin{document}

\begin{frontmatter}



\title{Probabilistic Forecasting of Real-Time Electricity Market Signals
\\ via Interpretable Generative AI\tnoteref{fund}}
\tnotetext[fund]{This work was supported in part by the National Science Foundation under Award EECS 2218110.}


\author[label1]{Xinyi Wang} 
\ead{xw555@cornell.edu}
\author[label1]{Qing Zhao}
\ead{qz16@cornell.edu}
\author[label1]{Lang Tong\corref{cor}}
\ead{lt35@cornell.edu}
\cortext[cor]{Corresponding author}
\affiliation[label1]{organization={School of Electrical and Computer Engineering},
            addressline={Cornell University}, 
            city={Ithaca},
            state={New York},
            postcode={14853}, 
            country={USA}}

\begin{abstract}
This paper introduces a generative AI approach to probabilistic forecasting of real-time electricity market signals, including locational marginal prices, interregional price spreads, and demand-supply imbalances. We present WIAE-GPF, a Weak Innovation AutoEncoder-based Generative Probabilistic Forecasting architecture that generates future samples of multivariate time series. Unlike traditional black-box models, WIAE-GPF offers interpretability through the Wiener-Kallianpur innovation representation for nonparametric time series, making it a nonparametric generalization of the Wiener/Kalman filter-based forecasting. A novel learning algorithm with structural convergence guarantees is proposed, ensuring that, under ideal training conditions, the generated forecast samples match the ground truth conditional probability distribution. Extensive tests using publicly available data from U.S. independent system operators under various point and probabilistic forecasting metrics demonstrate that WIAE-GPF consistently outperforms classical methods and cutting-edge machine learning techniques.
\end{abstract}



\end{frontmatter}




\section{Introduction}
\label{Sec:intro}
A key feature of generative AI is its ability to produce artificial samples that closely resemble real-world data. In particular, generative AI learns the underlying structure of a phenomenon from examples, enabling it to generate an arbitrarily large number of artificial samples that exhibit the same properties as the original data. Leveraging advanced neural networks and machine learning techniques, generative AI has achieved performance in real-world applications that far surpasses conventional methods \cite{Green23GenAI}.

The classical field of Probabilistic Forecasting (PF) naturally aligns with generative AI.  In particular, PF  predicts the conditional probability distribution of the future given past time series observations. Once this distribution is known, Monte Carlo samples of the future series can be generated. However, forecasting such conditional distributions presents significant computational and sampling challenges.  Specifically, nonparametric distribution forecasting is an infinite-dimensional functional estimation problem, often requiring finite-dimensional reductions, such as histograms with a finite number of bins or quantiles with a finite set of levels. Second, for continuous time series, each future realization is uniquely tied to a specific past history, making it inherently difficult to learn the true conditional distribution from data. The standard approach to PF is to assume a parametric model, reducing the infinite dimensional forecasting problem to finite dimensional prediction of distribution parameters. 

This paper applies generative AI principles to derive non-parametric Generative Probabilistic Forecasting (GPF), bypassing the obstacles of modeling, computational, and sample complexities associated with forecasting conditional probability distributions. Currently, no nonparametric GPF techniques exist.  By developing practical GPF solutions, we demonstrate the somewhat surprising conclusion that GPF problem is simpler and and more practical than PF.

As powerful as generative AI has become, it is often criticized as mysterious and uninterpretable, which casts doubt on whether the generated samples indeed follow the ground-truth distribution or merely appear to resemble the training data. For time series forecasting problems, it is difficult, if not impossible, to compare the similarities between training data and generated samples. No existing GPF techniques have established some level of guarantee the generative samples follow the correct conditional distributions. This paper aims to derive an {\em interpretable GPF} by making connections to classic theory of time series representation pioneered by Wiener, Kallianpur, and Rosenblatt \cite{Wiener:58Book,Rosenblatt:59}.
\subsection{Literature Review}
\label{subsec:review}
The literature on parametric and nonparametric probabilistic forecasting of electricity market signals is extensive. In this review, we focus on short-term (real-time) forecasting techniques for wholesale electricity prices and dispatch quantities, such as area demand-supply imbalances. Drawing on previous surveys \cite{hardle_review_1997,nowotarski_computing_2015,weron_forecasting_2008}, we place particular emphasis on machine learning-based techniques developed in the past decade.

Wholesale electricity prices and dispatch imbalances are endogenously determined by optimization-based market clearing processes, making them highly volatile due to their sensitivity to binding constraints. Consequently, real-time prices and dispatch quantities exhibit behavior distinct from exogenous physical processes like wind, solar, and demand time series. For this reason, we exclude the extensive body of energy forecasting literature, although those techniques can also be applied to price forecasting as well (for more on energy forecasting, see \cite{TaoPinsonEtal:20IEEEAcess} and references therein). Also excluded are forecasting techniques from the market operator's perspective, where network parameters and offers/bids are known. See, e.g., \cite{JiThomasTong17TPS} and references therein.

Probabilistic forecasting methods generally fall into parametric or nonparametric categories. Parametric approaches model future time series variables by predicting a parameterized conditional probability distribution, thus reducing an infinite-dimensional inference problem to a finite-dimensional. Popular parametric methods include autoregressive and moving average models \cite{ZhouEtal:IEEProc,GonzalezEtal:18TPS,UniejewskiWeron:21EE,LimaEtal_23TPS,chai_conditional_2019,SalinasEtal:19NeuripsDeepVAR}, Gaussian models \cite{dudek_multilayer_2016}, Student’s t-distribution \cite{LeeEtal:18TPS}, and others \cite{nowotarski_recent_2018}. While parametric models offer computational tractability, they often sacrifice accuracy due to model mismatches.

Nonparametric forecasting has a long history (see \cite{hardle_review_1997} for a review up to 1997). These methods estimate the underlying probability distribution or its properties, such as quantiles, without assuming a specific parametric form. However, classical nonparametric techniques \cite{Sheskin} face significant sample and computational challenges, especially when time series exhibit complex temporal dependencies.  Quantile regression is among the most popular techniques for forecasting electricity prices. By estimating multiple quantiles, it approximates the underlying probability density function through a histogram. A well-known application of quantile regression for day-ahead LMP forecasting can be found in \cite{uniejewski_regularized_2021}.

Over the past decade, deep learning technologies have significantly advanced point, PF, and GPF generative methods, utilizing various architectures and learning principles. Examples of these include Extreme Learning Machines (ELM) \cite{zhang_elm_2023}, Recurrent Neural Networks (RNN) \cite{ToubeauEtal_21TSG}, Variational Autoencoders (VAE) \cite{nguyen_temporal_2021,zheng_generative_2022,li_synergetic_2021,khodayar_convolutional_2020}, Long Short-Term Memory (LSTM) networks, diffusion models \cite{Li_2022_diffusion}, Generative Adversarial Network (GAN) \cite{ZhangMeng_22TPS} and Large Language Models (LLMs) featuring transformers and attention mechanisms \cite{BottieauEtal:23TPS}.

Among the deep-learning-based PF and GPF techniques, the VAE-based method \cite{nguyen_temporal_2021} is the most related to the proposed WIAE-GPF approach; both share a similar forecasting architecture, but differ in the training of the autoencoders. VAE typically relies on a parametric assumption, often on a conditional Gaussian distribution for the latent process, whereas WIAE-GPF imposes the innovation constraints that requires the latent process being independent and identically distributed and uniform (IID-uniform). See Sec.~\ref{Sec:repre}.

LLM-based generative AI, originally developed for linguistic time series, have recently demonstrated superior performance in various applications, including electricity price forecasting \cite{majumder2024exploring}. The transformer architecture, with its attention mechanism, has played a pivotal role in this success. However, despite their impressive results, these non-interpretable AI techniques give an understanding of the factors that would lead to effective probabilistic forecasting. In particular, there is no theoretical guarantee that LLM-based GPF can generate samples with the correct conditional probability distribution even when the training sample size is unbounded. Additionally, comprehensive empirical studies comparing LLMs to other machine learning methods for price forecasting remain scarce. To address this, we include three award-winning LLM-based models \cite{liu_pyraformer_2022,ZhouEtal:21AAAI} in our numerical comparisons in Section X, to benchmark their performance against other techniques.

\subsection{Summary of Contributions}
\begin{figure}[!h]
    \centering
    \includegraphics[width=0.75\linewidth]{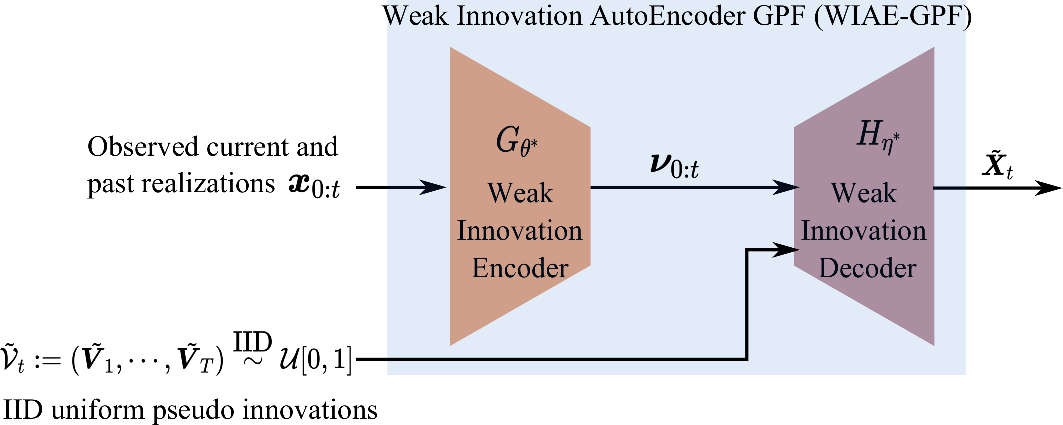}
    \caption{Forecasting pipeline for WIAE-GPF.}
    \label{fig:forecast}
\end{figure}

We propose Weak Innovation Autoencoder-based GPF (WIAE-GPF), a novel approach inspired by the classic Wiener-Kallianpur innovation representation of nonparametric time series \cite{Wiener:58Book} and a relaxation by Rosenblatt \cite{Rosenblatt:59}.  A key contribution of this work is to establish formally that the GPF architecture shown in Fig.~\ref{fig:forecast} is ``provable correct."   By provably correct,  we mean that with optimally trained WIAE autoencoder $(G_{\theta^*},H_{\eta^*})$, the WIAE-GPF output $\tilde{\Xbf}_{t}$ at time $t$ follows the probability distribution of the future variable $\Xbf_{t+T}$ given $(\Xbf_0=\xbf_0,\cdots, \Xbf_t=\xbf_t)$---the observed time series up to time $t$.  

WIAE-GPF stands out as an interpretable GPF model because of its connection to the classic Wiener/Kalman predictor under parametric Gaussian and state-space assumptions. Specifically, the Innovation Encoder in Fig.~\ref{fig:forecast} functions analogously to a causal whitening filter (via spectral factorization), while the Decoder corresponds to the linear Wiener predictor. Furthermore, the encoder's operation parallels the measurement update in Kalman filtering, and the decoder mirrors the prediction (time) update. In essence, WIAE-GPF can be viewed as a non-Gaussian nonparametric extension of the Wiener/Kalman predictor.

Note that the WIAE-GPF output $\tilde{\Xbf}_{t}$ is a function of observed time series $\xbf_{0:t}$ and independently  generated exogenous random vector $\tilde{\mathcal{V}}_t=(
\Vbf_1,\cdots, \Vbf_T)$ with IID uniformly distributed components, making $\tilde{\Xbf}_{t}$ a function of the random vector $\tilde{\mathcal{V}}_t$.  
By generating $\tilde{\Vc}_t$ from $T$ IID samples of uniform distribution, WIAE-GPF produces realizations of $\tilde{\Xbf}_t$ following the same conditional distribution as $\Xbf_{t+T}$. The formal definition of WIAE and its learning algorithm are presented in Sec.~\ref{Sec:repre}.  The WIAE-GPF architecture and its validity are shown in Sec.~\ref{Sec:GPF}.

Because practical implementations of WIAE are necessarily finite-dimensional, we establish a structural convergence property that the conditional distribution of the WIAE-GPF output converges to that of the conditional distribution of the time series. See Sec.~\ref{subsec:structural_conv} for details.

There have been some but limited applications of generative AI techniques in power system operations despite their accelerated advances in representing and learning time series models.  
Missing in particular are the validation and comparative studies using real-world market data.  
We fill this gap by comparing the WIAE-GPF with leading traditional and machine-learning algorithms for three applications: real-time LMP forecasting for energy markets, interregional LMP spread forecasting for interchange markets, and area control error (ACE) forecasting for regulation markets. 
Such comparisons are essential because these real-time market signals have characteristics not present in media signals such as video and natural language time series, where machine learning techniques have demonstrated success. 
Both LMP and ACE are highly dynamic time series with frequent spikes.  
Our comparison study offers a compelling case for WIAE-GPF across multiple performance measures for point and probabilistic forecasters.


The idea of WIAE-GPF was first presented in a preliminary version of this work \cite{WangTongZhao24CISS}, from which the current paper makes substantial new contributions\footnote{{Based on Turnitin comparison, this paper exhibits less than 15\% percent overall similarity and less than 4\% similarity to the preliminary version.}}.
In particular, the Bayesian sufficiency theorem in Sec.~\ref{subsec:forecasting} is significantly stronger than that in \cite{WangTongZhao24CISS}. 
Also new are a theorem (Theorem~\ref{thm:val} in Sec.~\ref{subsec:forecasting}) that establishes formally the validity of WIAE-GPF and a theorem on the structural convergence (Theorem~\ref{thm:converge} in Sec.~\ref{subsec:structural_conv}). 
We considered three specific real-time market applications, two of the three were not considered in \cite{WangTongZhao24CISS}.  
The numerical results for all three applications as well as the analysis and discussions are all new.

\subsection{Organization and Notations}
This paper is organized as follows. 
Sec.~\ref{Sec:repre} defines a nonparametric time series model, its innovation representations, and the learning algorithm of WIAE. 
Sec.~\ref{Sec:GPF} develops WIAE-GPF, the proposed GPF algorithm. 
Sec.~\ref{Sec:simulation} presents the application of WIAE-GPF in three market operations and the comparison studies of major forecasting benchmarks.

The notations used in this paper are standard. Random variables are in capital letters and their realizations in lowercases. 
Boldface letters are typically used for vectors and matrices.
We use $(\Xbf_t)$ to denote a multivariate random time series, where column vector $\Xbf_t =(X_{1t},\cdots, X_{dt})$ is the time series at time $t$, and $(X_{it})$ the $i$th sub-time series of $(\Xbf_t)$.
In this paper, $\Xbf_{t_1:t_2}$ denotes the segment of $(\Xbf_t)$ from $t_1$ to $t_2$, \ie $\Xbf_{t_1:t_2}=(\Xbf_{t_1}, \cdots, \Xbf_{t_2})$.
For two random vectors $\Xbf$ and $\Ybf$, $\Xbf\stackrel{\mbox{\tiny a.s.}}{=}\Ybf$ means the two random variables equal almost surely, and $\Xbf\stackrel{\mbox{\tiny d}}{=}\Ybf$ means the two equal in distribution.  
An IID random sequence with marginal distribution cumulative distribution $F$ is denoted by $\Xbf_t \stackrel{\mbox{\tiny\sf  IID}}{\sim} F$.  
Table~\ref{tab:notation} shows the major designated symbols used in the paper.
\begin{table}[htbp]
    \centering
    \caption{{\small Abbreviations and mathematical notations used in this paper.}}
    \vspace{0.5em}
    \label{tab:notation}
        \setlength\tabcolsep{1pt} 

    \resizebox{\linewidth}{!}{
    \begin{tabular}{l l}
    \hline
        GPF & Generative Probabilitic Forecasting.\\
        WIAE & Weak Innovation AutoEncoder.\\
        ACE & Area Control Error.\\
        LLM & Large Language Model.\\
        NMSE & Normalized Mean Square Error.\\
        NMAE & Normalized Mean Absolut Error.\\
        MASE & Mean Absolute Scaled Error.\\
        sMAPE & Symmetric Mean Absolute Percentage Error.\\
        CRPS & Continuous Ranked Probability Score.\\
        CP & Coverage Probability.\\
        CPE & Coverage Probability Error.\\
        NCW & Normalized Coverage Width.\\
        $(\Xbf_t)$ &The random process of predictive interests.\\
        $(\Vbf_t)$ &The innovation sequence. \\
        $(\Ubf_t)$ &An IID sequence of uniform distribution.\\
        $(\hat{\Xbf}_t)$ &The rescontruction sequence output by WIAE decoder.\\
        $\left(\hat{\Vbf}_{t}^{(m)}\right)$ &The weak innovation sequence estimated by a $m$-dimensional WIAE.\\
        $\left(\hat{\Xbf}_{t}^{(m)}\right)$ &The reconstruction sequence estimated by a $m$-dimensional WIAE.\\
        $(\xbf_t)$ &A sequence of real numbers indicating the past realizations of $(\Xbf_t)$.\\
        $(\nubf_t)$ &A sequence of real numbers indicating the past realizations of $(\Vbf_t)$.\\
        $G$ &WIAE encoder function.\\
        $H$ &WIAE decoder function.\\
        $G_\theta$ &A neural network approximation of $G$ parameterized by $\theta$.\\
        $H_\eta$ &A neural network approximation of $H$ parameterized by $\eta$.\\
        $D_\gamma$ &Innovation discriminator that measures the distance between $(\Vbf_t)$ and $(\Ubf_t)$.\\
        $D_\omega$  &Reconstruction discriminator that measures the distance between $(\Xbf_{0:t+T})$ and $(\Xbf_{0:t},\hat{\Xbf}_{t+1:t+T})$.\\
        $\Uc[0,1]^d$ &The continuous $d$-dimensional uniform distribution on $[0,1]$.\\
    \hline
    \end{tabular}}
    
\end{table}


\section{Innovation Representation Learning}
\label{Sec:repre}
\subsection{Strong and Weak Innovation Representations}
\label{subsec:inn-def}
\begin{figure}[htbp]
    \centering
    \includegraphics[width=0.5\linewidth]{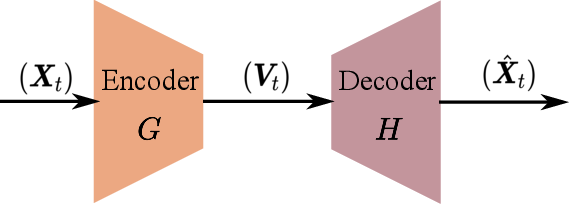}
    \caption{An autoencoder interpretation of innovation representations.}
    \label{fig:inn-ae}
\end{figure}

In 1958, Wiener and Kallianpur proposed an innovation representation of scalar time series \cite{Wiener:58Book}. 
In the parlance of modern machine learning, an innovation representation is a {\em causal autoencoder} shown in Fig.~\ref{fig:inn-ae} with the latent process $(\Vbf_t)$ being an IID-uniform {\em innovation sequence}.  
In particular, $\Vbf_t$ represents the new information (innovation) contained in $\Xbf_t$ independent of the past $\Xbf_{0:t-1}=(\Xbf_{t-1},\Xbf_{t-2},\cdots)$.  
Mathematically, the innovation representation of the time series is defined by causal mappings $(G,H)$ and $(\Vbf_t)$:
\begin{subequations}
\renewcommand{\theequation}{\theparentequation.\arabic{equation}}
    \begin{align}
    \Vbf_t &= G(\Xbf_t,\Xbf_{t-1},\cdots),\label{Eq:encoder}\\
    &(\Vbf_t)\stackrel{\mbox{\sf\tiny IID}}{\sim}\Uc[0,1]^d,\label{Eq:iid}\\
    \hat{\Xbf}_t &= H(\Vbf_t,\Vbf_{t-1},\cdots),\label{Eq:decoder}
\end{align}
\label{Eq:IAE}
\end{subequations}

The Wiener-Kallianpur's innovation autoencoder requires further that the decoder output $(\hat{\Xbf}_t)$ reconstructs the input $(\Xbf_t)$ (almost surely), \ie $(\hat{\Xbf_t})\stackrel{\mbox{\tiny a.s.}}{=} (\Xbf_t)$, which makes Wiener-Kallianpur's autoencoder a {\em strong innovation Autoencoder}.  
The perfect causal reconstruction implies that the innovation sequence $(\Vbf_t)$ is a sufficient statistic for all decision-making based on $(\Xbf_t)$. Therefore, using the IID-uniform $(\Vbf_t)$ for decision-making incurs no performance loss.  

However, Rosenblatt showed that the Wiener-Kallianpur (strong) innovation representation does not exist for broad classes of random processes, including some of the widely used finite-state Markov chains \cite{Rosenblatt:59}. Rosenblatt suggested a {\em weak innovation representation}, requiring that the autoencoder output $(\hat{\Xbf}_t)$ matches its input $(\Xbf_t)$ only in distribution:
\begin{align}
    (\Xbf_{0:t},\hat{\Xbf}_{t+1:t+T})\stackrel{\mbox{\tiny d}}{=} (\Xbf_{0:t+T}),\forall t.\label{Eq:recons}
\end{align}
Herein, we call the autoencoder $(G,H)$ for the weak innovation representation the Weak Innovation Auto Encoder (WIAE). 
\subsection{Innovation Representation Learning}
Beyond the Gaussian and additive Gaussian models, there is no known algorithm to obtain WIAE, especially when the underlying time series is nonparametric with an unknown probability structure. 
In \cite{WangTong:21JMLR}, the authors proposed a GAN-based learning of the strong innovation representation by jointly minimizing the Wasserstein distance of the latent process from the uniform IID process and the mean squared error ($l_2$ distance) of the autoencoder output as the estimate of the input. 
However, strong innovation representation applies only to a restricted class of time series, and the joint optimization of autoencoder with mixed Wasserstein and $l_2$ distance measures can be challenging.  
Finally, learning a scalar innovation representation limits the ability to incorporate multiple time series observations.  
The WIAE learning proposed below overcomes these shortcomings.
\subsection{WIAE Learning}
\label{subsec:training}
\begin{figure}[htbp]
    \centering
    \includegraphics[width=0.5\linewidth]{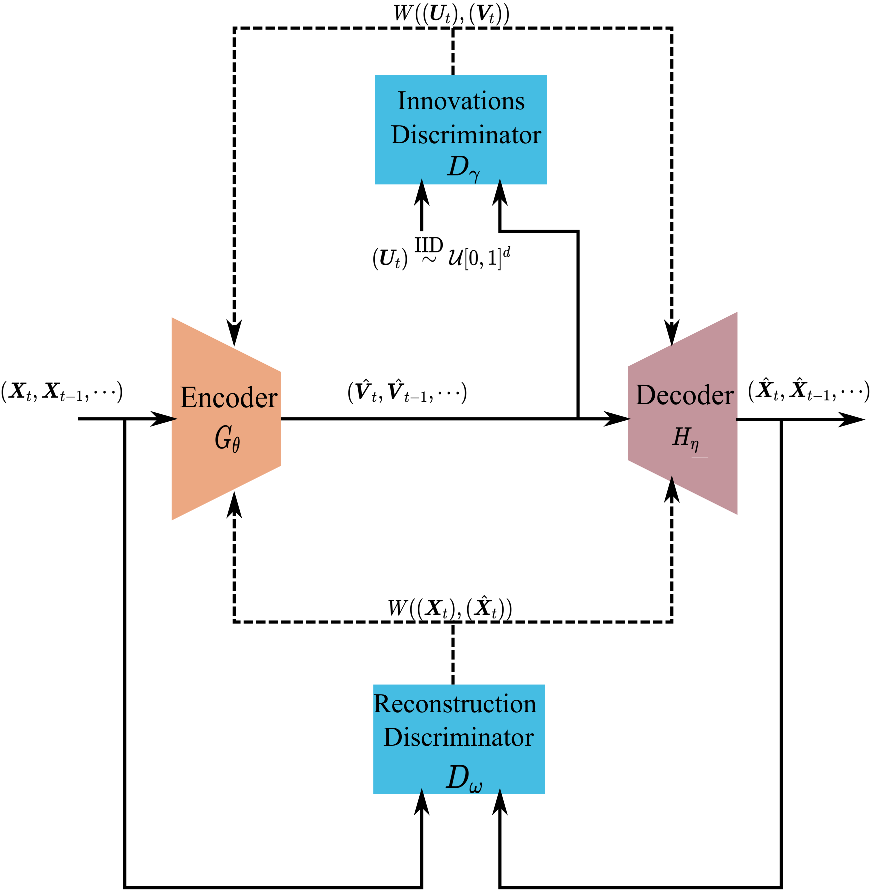}
    \caption{Training schematics of WIAE. Dash lines indicate the flow of training information.}
    \label{fig:schematic}
\end{figure}

We present a deep learning approach to learn a WIAE for the weak innovation representation defined in \eqref{Eq:recons}.
Shown in Fig.~\ref{fig:schematic} is the schematic that highlights key components of the WIAE learning.

The encoder $G_\theta$ and decoder $H_\eta$ are causal convolutional neural networks parameterized by coefficients $\theta$ and $\eta$, respectively.  
The weak innovation representation, at its core, matches the input-output distributions and constrains the latent process $(\Vbf_t)$ to be IID-uniform.
To this end, we introduce two neural network discriminators, the innovation discriminator $D_\gamma$ and the reconstruction discriminator $D_\omega$ with parameters $\gamma$ and $\omega$ respectively, to enforce \eqref{Eq:iid} and \eqref{Eq:recons}. 
In particular, the {innovation discriminator $D_\gamma$} compares the distributions of $(\hat{\Vbf}_t)$ and $(\Vbf_t)$, and the {reconstruction discriminator $D_\omega$} the compares joint distributions of $\Xbf_{0:t+T}$ and $(\Xbf_{0:t},\hat{\Xbf}_{t+1:t+T})$ with sufficiently large $T$.
These discriminators produce error signals to update neural network parameters $(\theta, \eta, \gamma, \omega)$.
{In this work, we adopt the Wasserstein discriminator proposed in \cite{Arjovsky17} to compute the Wasserstein distance between distributions.}

Because the two discriminators are both based on the Wasserstein-distance measure, their parameters $(\theta,\eta,\omega,\gamma)$ can be obtained via a single optimization:
\begin{multline}
     L:= \min_{\theta,\eta}\max_{\gamma,\omega}\big(\mbbE[D_{\gamma}\left((\Ubf_t)\right)] - \mbbE[D_\gamma((\hat{\Vbf}_{t}))] \\
     + \lambda(\mbbE[D_{\omega}(\Xbf_{0:t+T})]
      -\mbbE[D_{\omega}((\Xbf_{0:t},\hat{\Xbf}_{t+1:t+T}))])\big),
  \label{eq:loss}
\end{multline}
where $\lambda$ is a real number that scales the two Wasserstein distances.
The two parts of the inner maximization loop of loss function \eqref{eq:loss} regularize $(G_\theta,H_\eta)$ according to \eqref{Eq:iid} and \eqref{Eq:recons}.
It's evident that minimizing the inner loop with respect to $\theta$ and $\eta$ is equivalent to enforcing $(\Vbf_t)$ being IID uniform, and $(\Xbf_{0:t},\hat{\Xbf}_{t+1:t+T})$ having the same distribution as $\Xbf_{0:t+T}$. 
The training of the four neural networks is standard. Here we used the off-the-shelf Adam optimizer.

In a practical implementation of WIAE, finite (input) dimensional neural networks are used. The training of a finite-dimensional WIAE via \eqref{eq:loss} must also be implemented by finite segments of the random processes involved. 
In Sec.~\ref{subsec:structural_conv}, we consider the implications of such practical restrictions.

\section{WIAE-GPF and its Properties}
\label{Sec:GPF}
In this section, we introduce WIAE-GPF---a generative probabilistic forecasting techniques based on weak innovations representation.
Specifically, given past observations $\xbf_{0:t}$, WIAE-GPF produces (arbitrarily many) samples of $\tilde{\Xbf}_{t}$ that has the conditional distribution of $\Xbf_{t+T}$.
We present next the structure of WIAE-GPF, the Bayesian sufficiency of WIAE, and a structure convergence when WIAE is implemented with finite-dimensional implementations.

\subsection{Structure of WIAE-GPF}
\label{subsec:forecasting}
The structure of the proposed WIAE-GPF forecaster is shown in Fig.~\ref{fig:forecast}.  
At time $t$, given the realization of $\Xbf_{0:t}=\xbf_{0:t}$ and autoencoder $(G_{\theta^*},H_{\eta^*})$ trained by \eqref{eq:loss}, $\xbf_{0:t}$ up to time $t$, the WIAE encoder $G_{\theta^*}$ generates the innovation sequence $\nubf_{0:t}$.  
The WIAE decoder $H_{\eta^*}$ maps $\nubf_{0:t}$ and independently generated IID-uniform pseudo innovations $\tilde{\Vc}_{t} \stackrel{\mbox{\sf\tiny IID}}{\sim} \Uc[0,1]^T$ to produce a sample $\tilde{\Xbf}_{t}=\tilde{\xbf}_{t}$.   

Note that when forecasting $\Xbf_{t+T}$, we do not have realizations for random samples of $\Xbf_{t+1:t+T}$.   
The salient feature of WIAE-GPF is to replace samples from the unknown and arbitrarily distributed $\Xbf_{t+1:t+T}$ by realizations of {\em pseudo innovations} $\tilde{\mathcal{V}}_{t}$ known to be IID-uniform.
Thus, once the autoencoder is trained, generating random samples with the conditional distribution of $\Xbf_{t+T}$ is trivial. 

We now establish the validity of WIAE-GPF by showing that the WIAE-GPF output $\tilde{\Xbf}_{t}$ has the same conditional distribution as $\Xbf_{t+T}$ given $\Xbf_{0:t} = \xbf_{0:t}$.  
This is not obvious because the input of $H_{\eta^*}$ includes an exogenous random vector $\tilde{\Vc}_t$ and the weak innovation $(\Vbf_{0:t})$ that may not be a sufficient statistic.  

We first show that the weak innovation sequence is {\em Bayesian sufficient\footnote{$T(X)$ is Bayesian sufficient for the estimation of a random variable $Y$ if the posterior distribution of $Y$ given $X$ is the same as the one given $T(X)$ \cite{BickelDoksum07:book}.}}, which implies that any stochastic decision involving future time series $\Xbf_{t+T}$ can be made without loss based on the innovations $\Vbf_{0:t}$.   The same result was first presented in \cite{WangTongZhao24CISS} under the more restrictive setting of $H_{\eta^*}$ being injective.
\begin{lemma}[Bayesian Sufficiency of Multivariate Weak Innovations]
Let $(\Xbf_t)$ be a stationary time series for which the weak innovation representation exists. 
Let $(\Vbf_t)$ be the weak innovation representation of $(\Xbf_t)$. 
Then, for all $\xbf$ and $\Xbf_{0:t}=\xbf_{0:t}$, 
\begin{align}
    \Pr[\Xbf_{t+T}\leq\xbf|\Xbf_{0:t}=\xbf_{0:t}]
    =\Pr\left[\hat{\Xbf}_{t+T}\leq \xbf|\Vbf_{0:t}=\nubf_{0:t}\right].
\end{align}
\label{thm:sufficiency}
\end{lemma}
{\em Proof:} 
By the definition of weak innovation representation, 
\begin{equation}\label{eq:suf}
    \begin{aligned}
    \Pr[\Xbf_{t+T}\leq\xbf|\Xbf_{0:t}&=\xbf_{0:t}]
    =\Pr[\hat{\Xbf}_{t+T}\leq\xbf|\Xbf_{0:t}=\xbf_{0:t}]\\
    &\stackrel{(a)}{=}\Pr[\hat{\Xbf}_{t+T}\leq\xbf|G_{\theta^*}(\Xbf_{0:t})=G_{\theta^*}(\xbf_{0:t}))]\\
    &=\Pr[\hat{\Xbf}_{t+T}\leq\xbf|\Vbf_{0:t}=\nubf_{0:t}],\\
\end{aligned}
\end{equation}
where $(a)$ is from the Markovian structure of the autoencoder,\ie $\Xbf_{0:t}\stackrel{G_{\theta^*}}{\rightarrow}\hat{\Vbf}_{0:t}\stackrel{H_{\eta^*}}{\rightarrow}\hat{\Xbf}_{0:t}$.
By definition of Bayesian statistics \cite{BickelDoksum07:book},  $\Vbf_{0:t}=G_{\theta^*}({\Xbf_{0:t}})$ is a sufficient statistics for $\Xbf_{t+T}$ for all $T>0$.
$\square$

The validity of WIAE-GPF is shown next.

\begin{theorem}[Validity of WIAE-GPF]
For all $T>0$, the conditional distribution of the WIAE-GPF output $\tilde{\Xbf}_{t}$ given $\Xbf_{0:t}=\xbf_{0:t}$ is identical to that of $\Xbf_{t+T}$ (given $\Xbf_{0:t}=\xbf_{0:t}$), \ie,
\begin{align}
    \Pr[{\Xbf}_{t+T}\leq \xbf|\Xbf_{0:t}=\xbf_{0:t}]=\Pr[\tilde{\Xbf}_{t}\leq \xbf|\Xbf_{0:t}=\xbf_{0:t}].
    \label{eq:val_result}
\end{align}
\label{thm:val}
\end{theorem}
{\em Proof:} By Lemma~\ref{thm:sufficiency},
\begin{align}
    \Pr[\Xbf_{t+T}\leq \xbf|\Xbf_{0:t}=\xbf_{0:t}]=\Pr[\hat{\Xbf}_{t+T}\leq \xbf|\Vbf_{0:t}=\nubf_{0:t}],
    \label{eq:val_1}
\end{align}
where $\Vbf_{0:t}=G_{\theta^*}(\Xbf_{0:t})$ and $\nubf_{0:t}=G_{\theta^*}(\xbf_{0:t})$. Now consider
\[\begin{cases}
    \hat{\Xbf}_{t+T} = G_{\theta^*}(\Vbf_{0:t},\Vbf_{t+1:t+T})\\
    \tilde{\Xbf}_{t} = G_{\theta^*}(\Vbf_{0:t},\tilde{\Vc}_t),
\end{cases}\]
where, by definition, $(\Vbf_{0:t},\Vbf_{t+1:t+T},\tilde{\Vc}_{t})$ are jointly independent IID uniform sequences, and $\tilde{\Vc}_{t} \stackrel{\mbox{\tiny d}}{=} \Vbf_{t+1:t+T}$. Therefore, 
\begin{align}
    \Pr[\hat{\Xbf}_{t+T}\leq\xbf|\Vbf_{0:t}=\nubf_{0:t}]=
    \Pr[\tilde{\Xbf}_{t}\leq\xbf|\Vbf_{0:t}=\nubf_{0:t}].
    \label{eq:val_2}
\end{align}
Combining \eqref{eq:val_1} and \eqref{eq:val_2}, we have \eqref{eq:val_result}. $\square$

\subsection{Structural Convergence of WIAE-GPF}
\label{subsec:structural_conv}
This section focuses on the practical issue of finite-dimensional implementations of WIAE and discriminators in Fig.~\ref{fig:schematic}.
It is evident that no machine learning technique guarantees that a finite-dimensional implementation can extract weak innovations, even if the amount of historical samples available is unbounded.
Here we present a structural convergence result to show that, under the ideal training conditions with unbounded training samples and global convergence of training, the innovations generated from a finite-dimensional WIAE converge in distribution to the true weak innovations.

The structural convergence analysis assumes that the training samples are unbounded, and the training algorithm converges to the global minimum. 
Let $G_{\theta^*}^{(m)}$ be the optimally trained finite (input) dimensional CNN encoder that takes time-shifted $m$ consecutive observations $\Xbf_{t-m+1:t}$ and produces the latent process $\left(\hat{\Vbf}^{(m)}_t\right)$. 
Likewise, let $H^{(m)}_{\eta^*}$ be the optimally trained $m$-dimensional CNN decoder that produces the WIAE output sequence $\left(\hat{\Xbf}^{(m)}_t\right)$.   
Similarly defined are the finite dimensional discriminators that take $n$ consecutive inputs, denoted by $(D_{\omega}^{(n)},D_{\gamma}^{(n)})$.
In this paper, we choose $n=m$.

To analyze the asymptotic property of finite (input) dimensional WIAE-GPF, we make the following assumptions:
\ben
\item[A1] {\bf Existence:}  The random process $(\Xbf_t)$  has a weak innovation representation defined in (\ref{Eq:encoder} - \ref{Eq:decoder}) \& \eqref{Eq:recons}, and there exists a causal WIAE with continuous $G$ and $H$.
\item[A2] {\bf Feasibility:} There exists a sequence of finite input dimension auto-encoder functions {$(G_{\bar{\theta}}^{(m)}, H_{\bar{\eta}}^{(m)})$} that converges uniformly to $(G,H)$ under the mean-squared distance metric.
\item[A3] {\bf Training:} The training sample sizes are  infinite. The training algorithm for all finite-dimensional WIAE using finite-dimensional training samples converges almost surely to the global optimum.
\een
\begin{theorem} 
\label{thm:converge}
Under (A1-A3),  
\begin{align}
    (\Xbf_{0:t},\hat{\Xbf}_{t+T}^{(m)})\stackrel{\mbox{\tiny d}}{\rightarrow}(\Xbf_{0:t},\Xbf_{t+T}),~\forall t
\end{align}
as $m$ goes to infinity.
\end{theorem}
{\em Proof:} See \ref{subsec:conv_proof}.
 
\subsection{From GPF to Point and Quantile Forecasting}
GPF produces samples of the conditional probability distribution, from which point and quantile forecasts can be easily computed. 
Here we outline techniques to compute forecasts of several popular point and quantile forecasters.  To this end, let $\left\{\tilde{\xbf}_{t}^{(k)}, k=1,\cdots, K\right\}$ be the set of GPF generated samples from the probability distribution of the time series at time $t+T$ conditioned on past observations up to time $t$.
For the simplicity of mathematical expressions, we assume that $\left\{\tilde{\xbf}_{t}^{(k)}\right\}$ is sorted in the ascending order.

\bitem 
\item  {\bf Minimum Mean Squared Error (MMSE) Forecast:} The MMSE forecast is the mean of the conditional distribution.  The MMSE forecast $\hat{\xbf}^{\mbox{\tiny MMSE}}_{t}$  by a GPF is given by the conditional sample mean
\[
\hat{\xbf}^{\mbox{\tiny MMSE}}_{t} =\frac{1}{K} \sum_{k=1}^K \tilde{\xbf}_{t}^{(k)}.
\]
\item  {\bf Minimum Mean Absolute  Error (MMAE) Forecast:} The MMAE forecast is the median of the conditional distribution.   The MMAE forecast $\hat{\xbf}^{\mbox{\tiny MMAE}}_{t}$  by a GPF is given by the conditional sample median
\[
\hat{\xbf}^{\mbox{\tiny MMAE}}_{t} = \begin{cases}
    \tilde{\xbf}_{t}^{\left((K+1)/2\right)}, &\mbox{if $K$ is odd}\\
    0.5\left(\tilde{\xbf}_{t}^{\left(K/2\right)}+\tilde{\xbf}_{t}^{\left(K/2+1\right)}\right), &\mbox{if $K$ is even}.\\
\end{cases}
\]
\item {\bf Quantile Forecast:}  The GPF forecast of $q$-quantile $\hat{\xbf}^{\mbox{\tiny $q$-quantile}}_{t}$ is given by:
\[
\hat{\xbf}^{\mbox{\tiny $q$-quantile}}_{t} = \begin{cases}
    \tilde{\xbf}_{t}^{\left(qK\right)}, &\mbox{if $qK$ is an integer}\\
    0.5\left(\tilde{\xbf}_{t}^{\left([qK]\right)}+\tilde{\xbf}_{t}^{\left([qK]+1\right)}\right), &\mbox{otherwise},\\
\end{cases}
\]
where $[a]$ indicates the greatest integer not exceeding $a$.
\eitem

\subsection{Evaluation Metrics}
\label{subsec:metrics}
Comparing probabilistic forecasting methods is difficult due to the lack of ground truth for the underlying conditional distribution. 
However, because a GPF can produce arbitrarily many Monte Carlo samples\footnote{We used $1000$ Monte-Carlo samples and sample average to obtain point estimates.}, it can be evaluated by all point forecasting metrics.
More importantly, an ideal probabilistic forecaster that produces the correct conditional distribution will perform well under any point estimator metric under regularity conditions. 
Therefore, evaluating a GPF method based on a set of point-forecasting techniques is appropriate to assess its performance.
To this end, we also compared WIAE-GPF with some of the well-tested point forecasting techniques.

We used four popular point forecasting and two widely used probabilistic forecasting metrics. 
See~\ref{sec:metrics} for their definitions. 
Normalized mean squared error (NMSE) measures the error associated with the mean of the estimated conditional probability distribution.  Normalized Absolute Error (NMAE) measures the error associated with the median. 
Mean absolute scaled error (MASE) is the ratio of the NMAE of a method over that of the (naive) persistent predictor that uses the latest observation available as the forecast. 
Symmetric mean absolute percentage error (sMAPE) averages and symmetrizes the percentage error computed at each time stamp and is less sensible to outliers. 
The probabilistic forecasting metrics were the continuous ranked probability score (CRPS) \cite{TilmannEtal14ARStats}, Coverage Probability Error (CPE), and Normalized Coverage Width (NCW).
CRPS evaluates the quadratic difference between the predicted empirical cumulative density function (c.d.f.) with an indicator c.d.f. based on the ground truth.
CPE and NCW are often used to evaluate prediction intervals.
CPE is the deviation of the coverage probability (CP) from the nominal confidence level $\beta\%$, whereas NCW represents the width of the prediction intervals.
At similar level of CP, the method with smaller NCW shows better accuracy in prediction interval estimation.
In this paper, we computed the CPE and NCW of 10\%, 50\%, and 90\% intervals predicted by each probabilistic method.
The mathematical definition of those metrics can be found in the appendix.

For probabilistic forecasting techniques, we used their conditional means as the point forecasts when evaluated by NMSE, whereas the conditional median is used for NMAE, MASE, and sMAPE. For the quantile regression technique, BWGVT, we use the estimated 0.5-quantile as its point forecast for all metrics since it's unclear how to compute empirical mean from quantiles.
When producing interval forecasts for GPF methods, we took the empirical quantiles from the Monte-Carlo forecasts of future values.
In particular, the beta-coverage interval was defined as the $\beta$-width interval symmetric around the sample median.


\section{WIAE-GPF for Market Operations}
\label{Sec:simulation}
We now apply WIAE-GPF to forecasting market signals such as locational marginal prices and market imbalances. At the outset, we recognize that the underlying random processes are not known to be stationary, whereas WIAE-GPF is derived based on a representation of stationary processes. Here, we rely on the hypothesis that these processes are approximately stationary locally within the forecasting horizon. Our evaluations based on real market data presented here in some way validated this hypothesis. Brief discussions on the limitations and possible extensions can be found in Sec.~\ref{sec:conclusion}.

\begin{figure}[htbp]
\vspace{-1em}
    \centering
    \includegraphics[width=\linewidth]{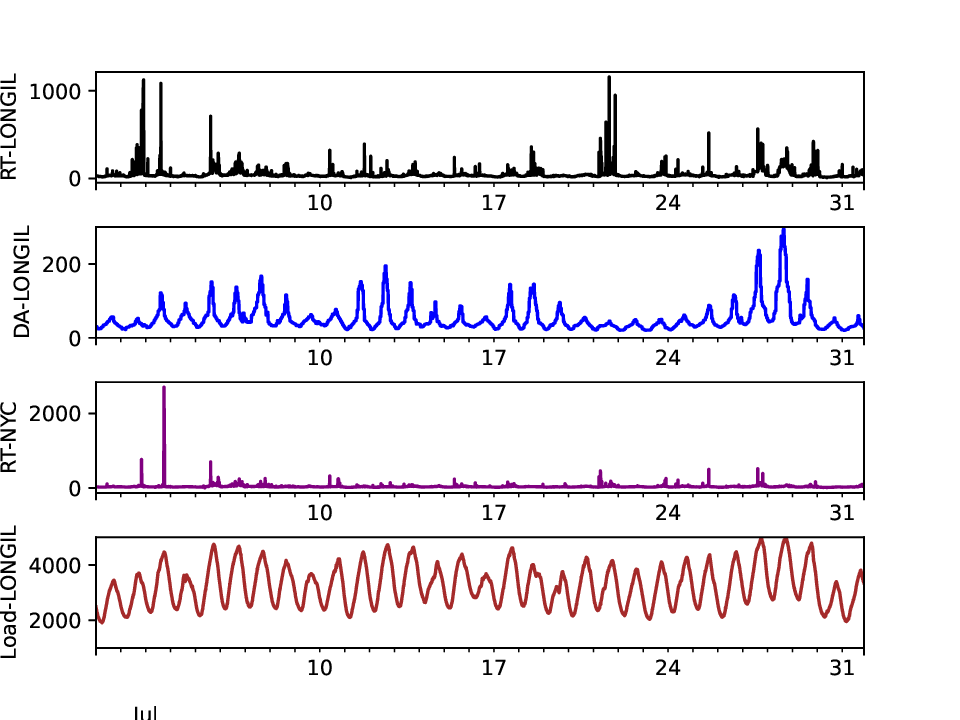}
    \caption{Real-time, day-ahead LMPs, and load at Long Island, and real-time LMP at NYC July 2023.}
    \label{fig:RTDA}
\end{figure}

We conducted extensive experiments to compare leading GPF and point forecasting techniques based on a suite of performance metrics. This section summarizes our findings for three market applications where GPF is particularly valuable to system operators and market participants: (a)  LMP forecasting for the optimal bidding in NYISO's 5-minute real-time energy market,  (b) GPF for the interregional LMP spread for the Coordinated Transaction Scheduling (CTS) \cite{Pike&White:21} market between NYISO and PJM, and (c) ACE forecasting for regulation services using PJM's 15-second ACE data.  Common to these applications is that the forecasted variables are endogenously determined by the market operations. In contrast to exogenous variables such as wind/solar generations and inelastic demands, the LMP and ACE values are the results of dispatch and commitment optimization, where binding constraints introduce spikes in dual variables from which LMPs are computed. They are highly dynamic as shown in Fig.~\ref{fig:RTDA}.

\vspace{-1em}
\subsection{Baseline Methods in Comparison}
\begin{table}[htbp]
    \centering
    \caption{{Comparison of the baselines.}}
    \vspace{0.5em}
    \resizebox{\linewidth}{!}{
    \begin{tabular}{|c|c|c|c|c|}
    \hline
        Algorithm&Forecasting Type &Time Series Model &Forecastor Output &ML Models \\
        \hline
        WIAE-GPF &Probabilistic &Nonparametric &Generative &CNN + WIAE\\
        \hline
        TLAE \cite{nguyen_temporal_2021} &Probabilistic &Parametric &Generative &RNN + VAE\\
        \hline              
        DeepVAR \cite{SalinasEtal:19NeuripsDeepVAR} &Probabilistic &Parametric (AR Model) &Model Parameters &LSTM\\
        \hline
        BWGVT \cite{BottieauEtal:23TPS}  &Probabilistic &Nonparametric &Forecasted Quantiles &LLM + Quantile Regression\\
        \hline
        Pyraformer \cite{liu_pyraformer_2022} &Point &Nonparametric  &Point Estimate &LLM\\
        \hline
        Informer \cite{ZhouEtal:21AAAI} &Point &Nonparametric  &Point Estimate &LLM\\
        \hline
    \end{tabular}}
    \label{tab:comparison}
\end{table}
We compared WIAE-GPF with six leading forecasters based on their relevance to power system applications and their established reputations. See Table \ref{tab:comparison} for attributes of these techniques and references. WIAE-GPF is the only nonparametric GPF forecaster. Because there are limited nonparametric GPF techniques, we also included in our comparison popular machine-learning-based parameterized GPF and point-forecasting techniques.

For deep-learning techniques, we compared with DeepVAR \cite{SalinasEtal:19NeuripsDeepVAR}, a multivariate generalization of the popular DeepVAR that has become a key baseline for time series forecasting in multiple applications.
Temporal Latent AutoEncoder (TLAE) \cite{nguyen_temporal_2021} is an autoencoder-based parametric GPF where the conditional distribution of the future time series variables is obtained by passing a conditional Gaussian process through the decoder, where the conditional mean and variance are computed by the encoder.  Once these parameters are estimated from observed realizations, Gaussian Monte Carlo samples are fed into the decoder to generate samples of the forecasted random variable. 

We also included three popular forecasting techniques based on LLMs.  One is the award-winning technique Informer \cite{ZhouEtal:21AAAI}; the other is  Pyraformer \cite{liu_pyraformer_2022} that captures temporal dependency at multiple granularities\footnote{Pyraformer is the keynote presentation at The International Conference on Learning Representations (ICLR) 2022.}.  
Pyraformer showed superior performance over a wide range of LLM-based point forecasting techniques.
Specifically developed for LMP forecasting,  BWGVT\footnote{We have named the algorithm by the first letters of authors' last names}
\cite{BottieauEtal:23TPS} combines quantile regression with a transformer architecture derived for LLMs.  

\subsection{LMP forecasting for Energy Market Participation}
\label{subsec:RTDA}
For a self-scheduled resource submitting a quantity bid to the energy market, the ability to forecast future prices is essential in constructing its bids and offers. 
With GPF generating future LMP realizations, the problem of optimal offer/bid strategies can be formulated as scenario-based stochastic optimization \cite{Tomasson&Hesamzadeh&Wolak:20AE}. 
Our experiment was based on a use case of a merchant storage owner submitting quantity offers and bids to a deregulated wholesale market, using LMP from NYISO as the hypothetical price realizations.

The real-time market of NYISO closes sixty minutes ahead of actual delivery, which means that the forecasting horizon needs to be longer than 60 minutes. 
Its real-time LMPs and load were collected every 5 minutes and day-ahead LMPs every hour.
Two experiments were conducted to produce probabilistic forecasts of 60-minute ahead LMPs at the Long Island (LONGIL) using (a) the day-ahead prices and the current and past real-time LMPs at LONGIL, along with the system load up to the time of submitting the bid; (b) the neighboring NYC real-time LMP in addition to the data in (a). 

Electricity prices are seasonal.
We selected July 2023, October 2023, January 2024, and April 2024 to represent summer, fall, winter, and spring, respectively. Our method, along with all baseline methods in Table.~\ref{tab:comparison}, were tested using NYISO real-time price data at LONGIL collected during these months. For every week in the selected months, a forecastor was trained using historical data from the preceding 30-day period. The same forecastor was used to produce forecast for that week.

Fig.~\ref{fig:RTDA} shows the real-time LMP trajectories at both LONGIL and NYC in July 2023, along with the demand and the day-ahead LMP at LONGIL.  
{The real-tine LMP at LONGIL exhibited the highest level of volatility for all four months.}
The real-time LMPs at LONGIL and NYC showed apparent spatial dependencies, while the dependency between day-ahead and real-time LMPs at LONGIL. 
The dependencies between load and real-time LMP were less obvious.
{The real-time LMPs for the other three seasons show similarities to those in summer, with fall and summer exhibiting the most volatility, while spring is the least volatile. During summer and fall, real-time LMPs frequently spiked above 1000\$/MWh due to high demand and weather-related factors. In contrast, winter and spring experienced more stable prices; though on April 29th at 7:40 PM, the real-time LMP at LONGIL surged to 6752\$/MWh before dropping back below 1000\$/MWh by 8 PM. }

\begin{table}[htbp]
    \centering
    \caption{Comparison of volatility between seasons}
    \vspace{0.5em}
    \resizebox{\linewidth}{!}{
    \begin{tabular}{|c|c|c|c|c|c|}
    \hline
         Season & Summer & Fall & Winter & Spring & Spring (excl. Apr. 29th, 2024)  \\
         \hline
         Standard Deviation &52.8266 &68.3016 &43.6466 & 96.8191 &21.6991 \\
         \hline
    \end{tabular}}
    \label{tab:season}
\end{table}

{The standard deviation for LMPs of each seasons are included in Table.~\ref{tab:season}. Although spring LMPs appear to have the highest volatility, the extreme outlier price (over 6500\$/MWh) on Apr. 29th, 2024 mainly contributes to the high standard deviation. Excluding the extreme price from the dataset, spring LMPs exhibited the least level of volatility. Winter volatility is higher than that of spring time (excluding Apr. 29th), though it is still lower than that of Summer and Fall.}

{As illustrated by Fig.~\ref{fig:RTDA}, the real-time LMPs at LONGIL exhibit significant spikes with absolute values far exceeding the rest. These spikes are challenging to predict, and even minor prediction errors for these spikes can disproportionately impact the overall evaluation metrics.
To ensure a more informative comparison among techniques, we calculated the evaluation metrics only on the ``normal" LMPs, defined as those within three standard deviations from the mean."}

\paragraph{\underline{Real-time LMP forecasting performance: an overall summary}}
{We analyzed the overall performance of all methods across the four seasons. WIAE-GPF consistently achieved the best overall performance on all evaluation metrics, both point and probabilistic. It ranked first in NMSE, CRPS, and CPE for nearly all cases, while it placed second in NMAE, MASE, and sMAPE in most instances. The strong performance of WIAE-GPF can be attributed to its focus on matching the conditional distribution, with a validity guarantee ensuring that the Monte Carlo samples generated have the same conditional distribution as the actual time series variable.
The strong performance of WIAE-GPF compared with benchmark techniques under NMSE, NMAE, and other point forecasting metrics suggests that the WIAE-GPF approximates the true conditional probability distribution well.}

{The second best-performing technique overall was DeepVAR, which excelled in NMAE and MASE for summer and fall. This indicates that DeepVAR achieved accurate parameter estimation for the mean and median of the Gaussian AR model it assumes. However, its CPE and CRPS metrics are generally worse than those of WIAE-GPF, possibly due to a model mismatch between the volatile and complex conditional distribution of real-time LMP and the Gaussian AR assumption of DeepVAR.}

{The three LLM-based estimators (BWGVT, Pyraformer, and Informer) performed similarly, outperforming TLAE but generally falling short of WIAE-GPF and DeepVAR. Pyraformer and Informer, both point forecasters trained to minimize mean squared forecasting error, performed better under NMSE. Notably, Pyraformer achieved the best NMSE among all methods during the winter month. However, their NMAE, MASE, and sMAPE scores were generally worse than those of WIAE-GPF and DeepVAR in most cases. Informer underperformed compared to Pyraformer across the other seasons, but it did achieve the best NMAE and MASE during the spring month, when LMP exhibited substantially lower volatility.
We conclude the underperformance of the LLM-based point estimators to the un-necessity of adopting attention mechanism for long-range dependency and the difficulty of training due to their large numbers of parameters. For details, see Sec.~\ref{subsec:discussion}.}

{The LLM-based probabilistic forecasting technique BWGVT uses quantile regression to predict future distributions. Its quantile predictions were sensitive to outliers, especially compared to the GPF methods that rely on a stochastic latent process, leading to poorer point forecasting performance. BWGVT tended to predict larger intervals, as indicated by its high NCW across all cases. As a result, its CPEs were always positive, meaning that the prediction intervals it generated covered more than the nominal percentages. Consequently, both its point estimation results and CRPS were worse than those of DeepVAR and WIAE-GPF.}

{TLAE performed the worst across the broad. Its point evaluation metrics consistently ranked the worst over all methods, while its probabilistic evaluation metrics are slighly better.
TLAE is a VAE-based auto encoder with a correlated Gaussian latent process. 
It generates samples of forecasts at the next time stamp by re-parameterization.
For a multiple timestamp-ahead prediction, TLAE drawns sample iteratively by substituting the mean of the samples generated for the previous timestamp.
This heuristic has no guarantee to match the conditional distribution of the generated samples and the conditional distribution of the future given the current observations, and it could also cause an accumulation of biases.  
}
\begin{landscape}
\begin{table*}[htbp]
    \centering
    \caption{{\small Performance evaluation of forecasting results for real-time price forecasting at Long Island. The numbers in the parentheses were the ranking of the algorithm.  
The columns under the label of LONGIL are the GPF performance of the 12-step foresting of the LMP at LONGIL. }}
\vspace{0.5em}
   \resizebox{\linewidth}{!}{
    \begin{tabular}{|c|c|c|c|c|c|c|c|c|c|c|}
    \hline
         Season &{Methods} & {NMSE}  & {NMAE}  &{MASE}  &{sMAPE} &  CRPS &CPE (90\%) [NCW] &CPE (50\%) [NCW] &CPE (10\%) [NCW]
        \\
        \hline\hline
         \multirow{6}{*}{\makecell{Summer\\
         (2023.07/01-2023.07.31)}}&WIAE-GPF  &(1) $\mathbf{0.1084}$   & (2) $0.2341$ &(2) $0.5812$ &(2) $0.2644$ & (1) $\mathbf{13.2607}$ &(1) $\mathbf{0.0009}[0.0718]$ &(1) $\mathbf{-0.0264}[0.0261]$ & (1) $\mathbf{0.0059}[0.0055]$
         \\
         \cline{2-10}
         &TLAE \cite{nguyen_temporal_2021}  &(6) $0.4690$ &(6) $0.4616$ &(6) $0.9768$  &(6) $0.4436$ &(4) $23.3288$ &(4) $-0.1034[0.0723]$ &(3) $0.1492[0.0268]$ &(3) $-0.0646[0.0027]$ 
         \\
         \cline{2-10}
         &DeepVAR \cite{SalinasEtal:19NeuripsDeepVAR} &(3) $0.2009$ &(1) $\mathbf{0.1849}$ &(1) $\mathbf{0.3911}$ &(1) $\mathbf{0.2496}$ &(2) $14.0716$ &(2) $0.0275[0.0777]$ &(2) $-0.0459[0.226]$ &(2) $-0.0243[0.0034]$         \\
         \cline{2-10}
        &BWGVT \cite{BottieauEtal:23TPS}  &(4) $0.2206$ &(4) $0.2804$ &(4) $0.5934$ &(3) $0.2711$ &(3) $15.0476$ &(3) $0.0968[0.0862]$ &(4) $0.1646[0.0419]$ &(4) $0.1279[0.0081]$
        \\
          \cline{2-10}
         &Pyraformer \cite{liu_pyraformer_2022}	&(2) $0.1090$ &(3) $0.2717$ &(3) $0.5749$ &(4) $0.2934$ &N/A &N/A &N/A &N/A
         \\
         \cline{2-10}
         &Informer \cite{ZhouEtal:21AAAI}    &(5) $0.4505$ &(5) $0.3935$ &(5) $0.8326$ &(5) $0.3329$ &N/A &N/A &N/A &N/A
         \\
        \hline\hline
         \multirow{6}{*}{\makecell{Fall\\(2023.10.01-2023.10.31)}} 
         &WIAE-GPF  &(1) $\mathbf{0.1020}$   & (2) $0.2806$ &(2) $0.4131$ &(2) $0.2522$ & (1) $\mathbf{10.9362}$ &(1) $\mathbf{0.0002}[0.0711]$ &(1) $\mathbf{-0.0390}[0.0292]$ & (1) $\mathbf{-0.0157}[0.0054]$
         \\
         \cline{2-10}
         &TLAE \cite{nguyen_temporal_2021}  &(6) $0.6181$ &(6) $0.4693$ &(6) $0.6908$  &(6) $0.4349$ &(4) $19.1394$ &(3) $-0.0487[0.0335]$ &(3) $-0.0395[0.0135]$ &(2) $0.0230[0.0059]$ 
         \\
         \cline{2-10}
         &DeepVAR \cite{SalinasEtal:19NeuripsDeepVAR} &(5) $0.5684$ &(1) $\mathbf{0.2328}$ &(1) $\mathbf{0.3427}$ &(3) $0.2662$ &(2) $12.6109$ &(2) $0.0205[0.0414]$ &(2) $-0.0328[0.106]$ &(3) $-0.0246[0.0018]$         \\
         \cline{2-10}
        &BWGVT \cite{BottieauEtal:23TPS}  &(4) $0.2992$ &(5) $0.3696$ &(5) $0.5441$ &(5) $0.3098$ &(3) $14.1423$ &(4) $0.0905[0.0837]$ &(4) $0.1763[0.0393]$ &(4) $0.0367[0.0074]$
        \\
          \cline{2-10}
         &Pyraformer \cite{liu_pyraformer_2022}	&(2) $0.1080$ &(3) $0.2949$ &(3) $0.4341$ &(1) $\mathbf{0.2116}$ &N/A &N/A &N/A &N/A
         \\
         \cline{2-10}
         &Informer \cite{ZhouEtal:21AAAI}    &(3) $0.2615$ &(4) $0.2985$ &(4) $0.4394$ &(4) $0.2823$ &N/A &N/A &N/A &N/A
         \\
        \hline
        \hline
        \multirow{6}{*}{\makecell{Winter\\(2024.01.01-2024.01.31)}} &WIAE-GPF  &(2) ${0.0906}$   & (2) $0.2263$ &(2) $0.3369$ &(1) $\mathbf{0.2307}$ & (1) $\mathbf{12.1949}$ &(1) $\mathbf{0.0014}[0.1203]$ &(2) ${0.0358}[0.0493]$ & (2) ${0.0196}[0.0092]$
         \\
         \cline{2-10}
         &TLAE \cite{nguyen_temporal_2021}  &(6) $0.2244$ &(6) $0.3306$ &(6) $0.4922$  &(6) $0.4355$ &(4) $20.7468$ &(3) $-0.0337[0.0371]$ &(3) ${-0.0369}[0.0159]$ &(4) $-0.0502[0.0027]$ 
         \\
         \cline{2-10}
         &DeepVAR \cite{SalinasEtal:19NeuripsDeepVAR} &(3) $0.1507$ &(1) $\mathbf{0.1614}$ &(1) $\mathbf{0.3273}$ &(2) $0.2499$ &(2) $13.1403$ &(2) $0.0211[0.1525]$ &(1) $\mathbf{-0.0160}[0.220]$ &(3) $-0.0287[0.0049]$         \\
         \cline{2-10}
        &BWGVT \cite{BottieauEtal:23TPS}  &(5) $0.1649$ &(4) $0.2725$ &(4) $0.4057$ &(3) $0.2515$ &(3) $15.2919$ &(4) $0.081[0.1808]$ &(4) $0.1804[0.0545]$ &(1) $\mathbf{0.0158}[0.0058]$
        \\
          \cline{2-10}
         &Pyraformer \cite{liu_pyraformer_2022}	&(1) $\mathbf{0.0721}$ &(5) $0.2786$ &(5) $0.4144$ &(5) $0.2747$ &N/A &N/A &N/A &N/A
         \\
         \cline{2-10}
         &Informer \cite{ZhouEtal:21AAAI}    &(4) $0.1553$ &(3) $0.2528$ &(3) $0.3763$ &(4) $0.2617$ &N/A &N/A &N/A &N/A
         \\
         \hline
         \hline
         \multirow{6}{*}{Spring (2024.04.01-2024.04.30)} &WIAE-GPF  &(1) $\mathbf{0.0797}$   & (2) $0.2137$ &(2) $0.2942$ &(2) ${0.1289}$ & (1) $\mathbf{5.3111}$ &(1) $\mathbf{0.0068}[0.1485]$ &(1) $\mathbf{0.0208}[0.0624]$ & (1) $\mathbf{0.0148}[0.0113]$
         \\
         \cline{2-10}
         &TLAE \cite{nguyen_temporal_2021}  &(6) $0.1903$ &(6) $0.2729$ &(6) $0.3756$  &(6) $0.4016$ &(4) $15.3330$ &(3) $-0.0510[0.1270]$ &(2) ${0.0273}[0.0635]$ &(2) $0.0217[0.0143]$ 
         \\
         \cline{2-10}
         &DeepVAR \cite{SalinasEtal:19NeuripsDeepVAR} &(5) $0.1105$ &(4) ${0.2345}$ &(4) ${0.3228}$ &(1) $\mathbf{0.1233}$ &(2) $5.9420$ &(2) $0.0154[0.1665]$ &(3) $-0.0423[0.572]$ &(3) $-0.0415[0.0069]$         \\
         \cline{2-10}
        &BWGVT \cite{BottieauEtal:23TPS}  &(3) $0.0877$ &(3) $0.2188$ &(3) $0.3011$ &(5) $0.2166$ &(3) $14.6103$ &(4) $0.0805[0.1551]$ &(4) $0.2436[0.0871]$ &(4) ${0.1177}[0.0196]$
        \\
          \cline{2-10}
         &Pyraformer \cite{liu_pyraformer_2022}	&(4) $0.1022$ &(5) $0.2890$ &(5) $0.3978$ &(3) $0.1788$ &N/A &N/A &N/A &N/A
         \\
         \cline{2-10}
         &Informer \cite{ZhouEtal:21AAAI}    &(2) $0.0845$ &(1) $\mathbf{0.1867}$ &(1) $\mathbf{0.2570}$ &(4) $0.1880$ &N/A &N/A &N/A &N/A
         \\
         \hline
    \end{tabular}}
    \label{tab:RTDA}
\end{table*}
\end{landscape}

\newpage
\paragraph{\underline{Summer real-time LMP prediction}}
{Test results for July 2023 are shown in row 2-7 of Table \ref{tab:RTDA} with the best performances highlighted in bold.  
We observed that WIAE-GPF performed the best for the probabilistic evluation metrics and NMSE, and ranked in close second for NMAE, MASE and sMAPE, whereas DeepVAR performed the best for those three metrics associated with conditional median.
Pyraformer, trained to minimize MSE, achieved the second-best NMSE but didn't achieve notable performance for the other metrics.
We observed that WIAE-GPF outperformed other methods in probabilistic evaluation metrics and NMSE, while ranking a close second in NMAE, MASE, and sMAPE. In contrast, DeepVAR excelled in these three metrics, which are associated with the conditional median.
Pyraformer, which was trained to minimize MSE, achieved the second-best NMSE but did not show notable performance in the other metrics.
We believe this may be due to DeepVAR, as a Gaussian AR process with fewer parameters and a simpler training objective, could perform better at estimating the conditional median.
On the other hand, WIAE-GPF, being nonparametric and trained with two Wasserstein discriminators, is more sensitive to hyperparameters and initialization. However, its nonparametric nature allows it to avoid the model misspecification issues that can affect parametric methods like DeepVAR when predicting the overall conditional distribution, as evidenced by its superior performance in CRPS and CPE.}
\begin{landscape}
    \begin{figure}
    \centering
    \begin{subfigure}[t]{0.24\linewidth}
        \includegraphics[width=\linewidth]{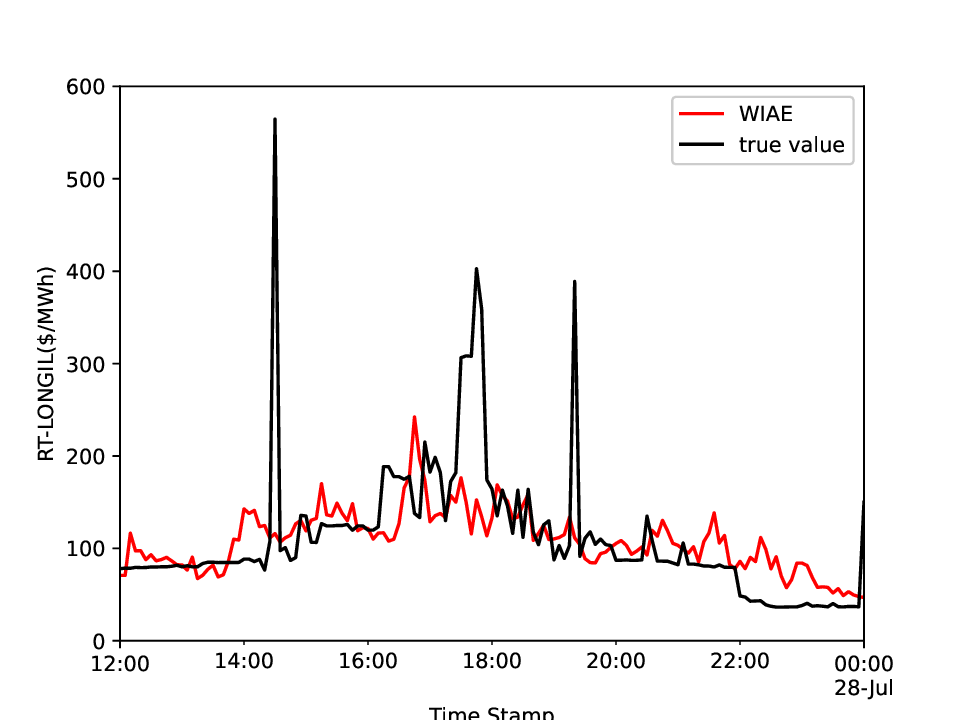}
        \caption{WIAE-GPF}
    \end{subfigure}
    \begin{subfigure}[t]{0.24\linewidth}
        \includegraphics[width=\linewidth]{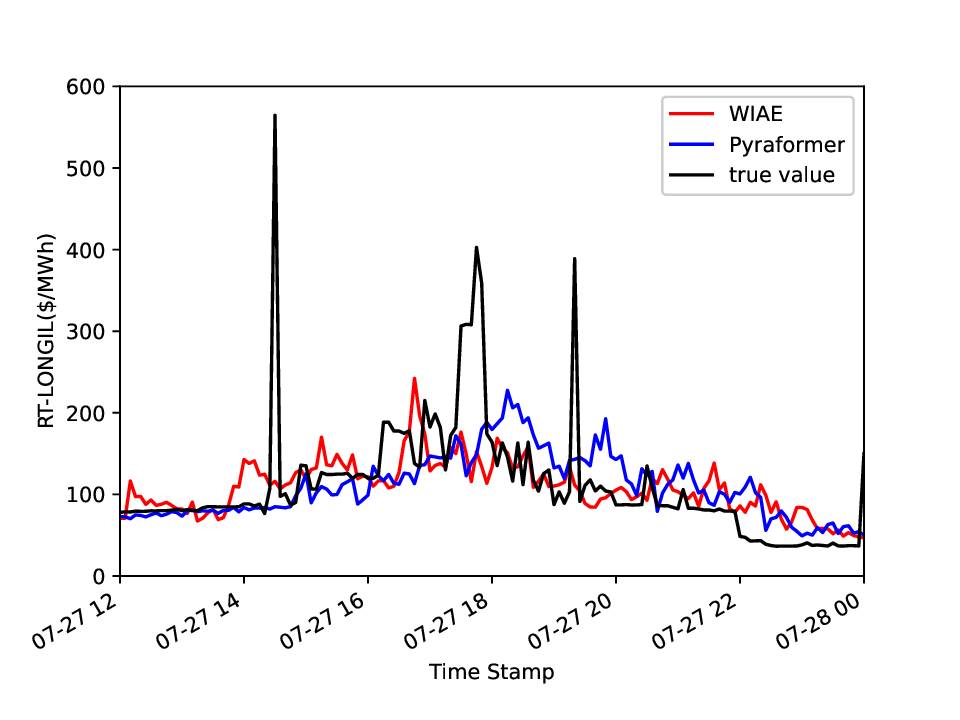}
        \caption{Pyraformer}
    \end{subfigure}
    \begin{subfigure}[t]{0.24\linewidth}
        \includegraphics[width=\linewidth]{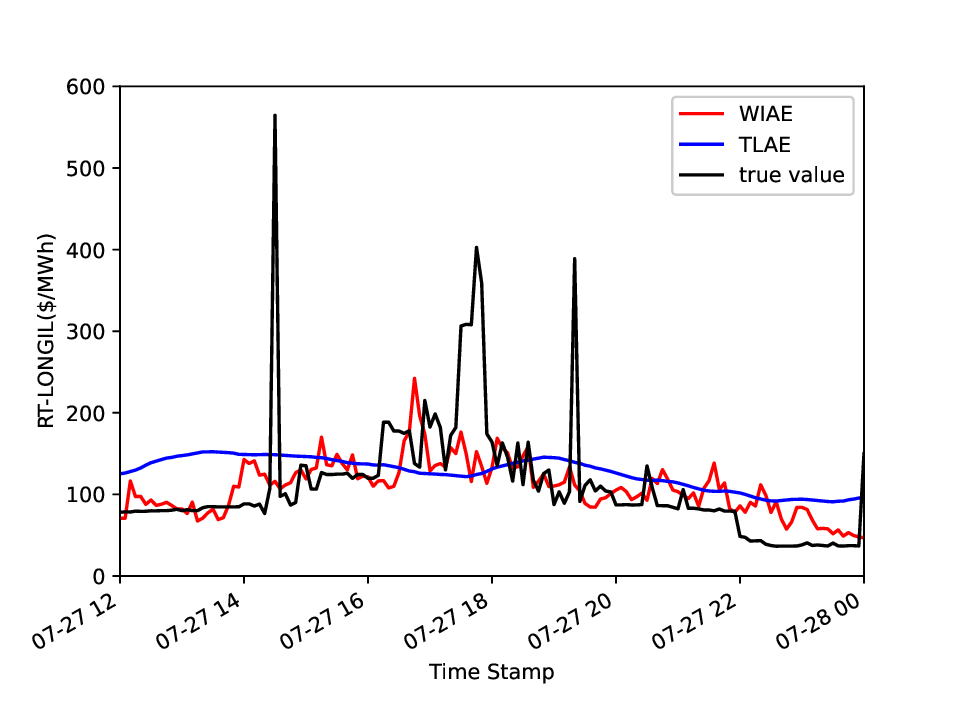}
        \caption{TLAE}
    \end{subfigure}
    \begin{subfigure}[t]{0.24\linewidth}
        \includegraphics[width=\linewidth]{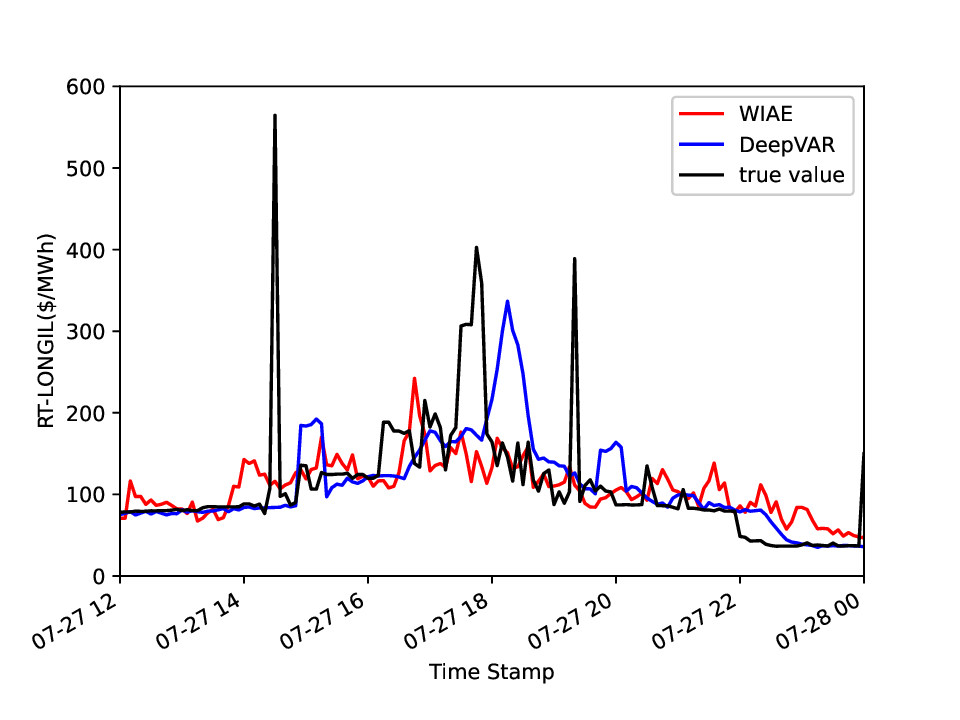}
        \caption{DeepVAR}
    \end{subfigure}
    \\
    \begin{subfigure}[t]{0.24\linewidth}
        \includegraphics[width=\linewidth]{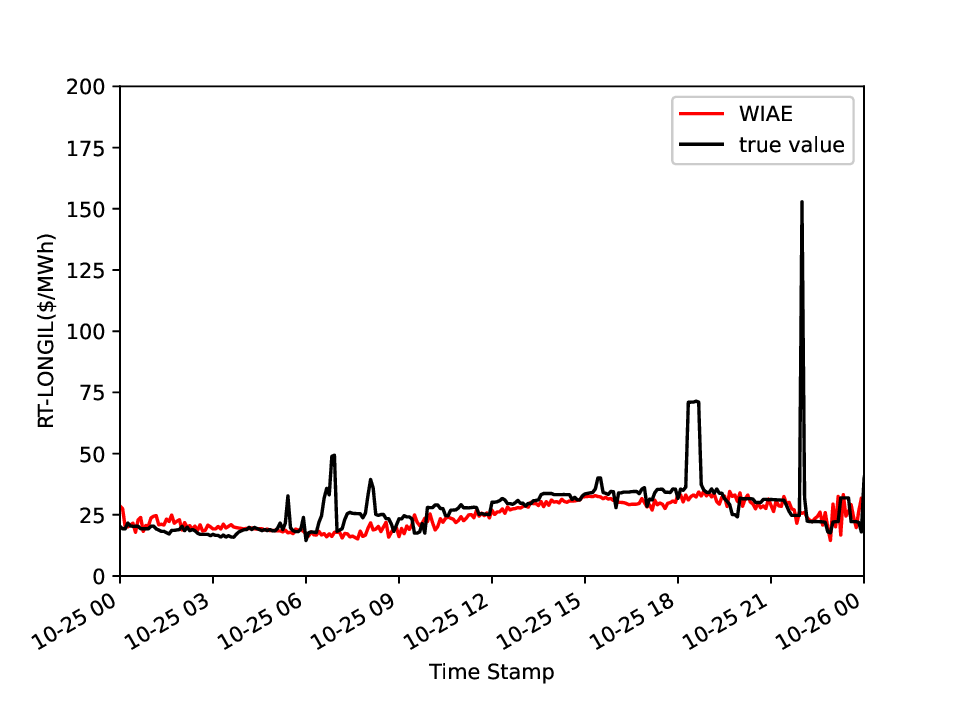}
        \caption{WIAE-GPF}
    \end{subfigure}
    \begin{subfigure}[t]{0.24\linewidth}
        \includegraphics[width=\linewidth]{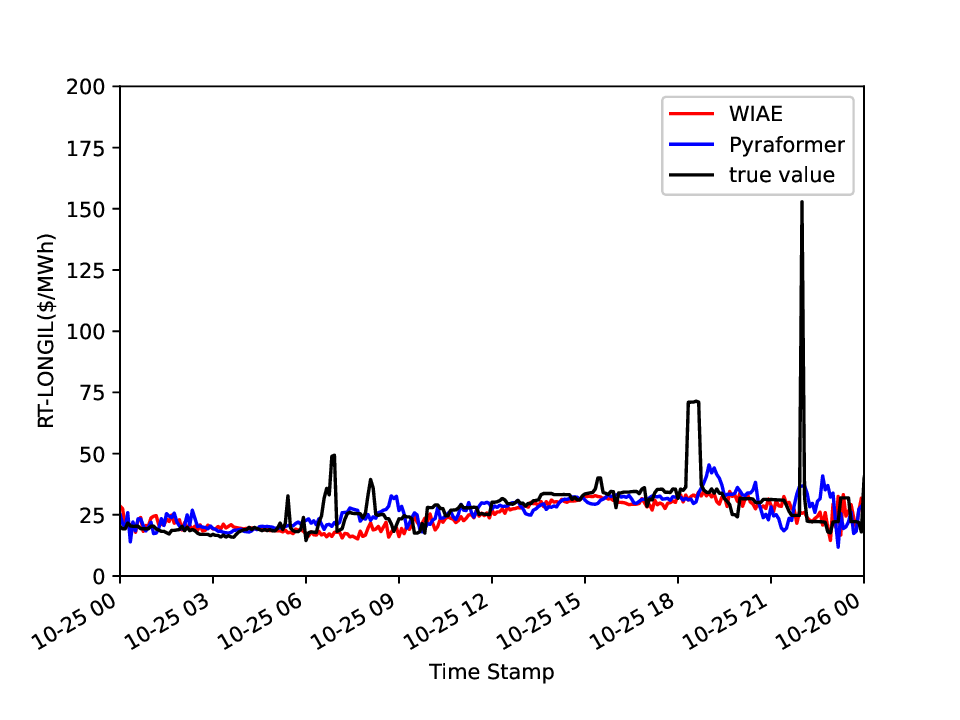}
        \caption{Pyraformer}
    \end{subfigure}
    \begin{subfigure}[t]{0.24\linewidth}
        \includegraphics[width=\linewidth]{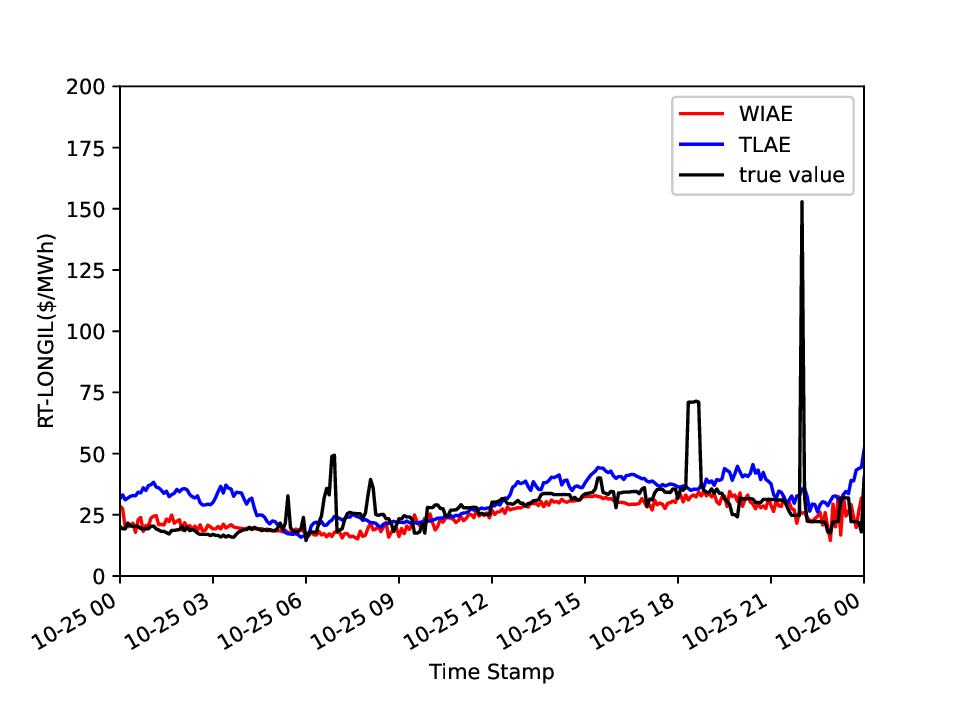}
        \caption{TLAE}
    \end{subfigure}
    \begin{subfigure}[t]{0.24\linewidth}
        \includegraphics[width=\linewidth]{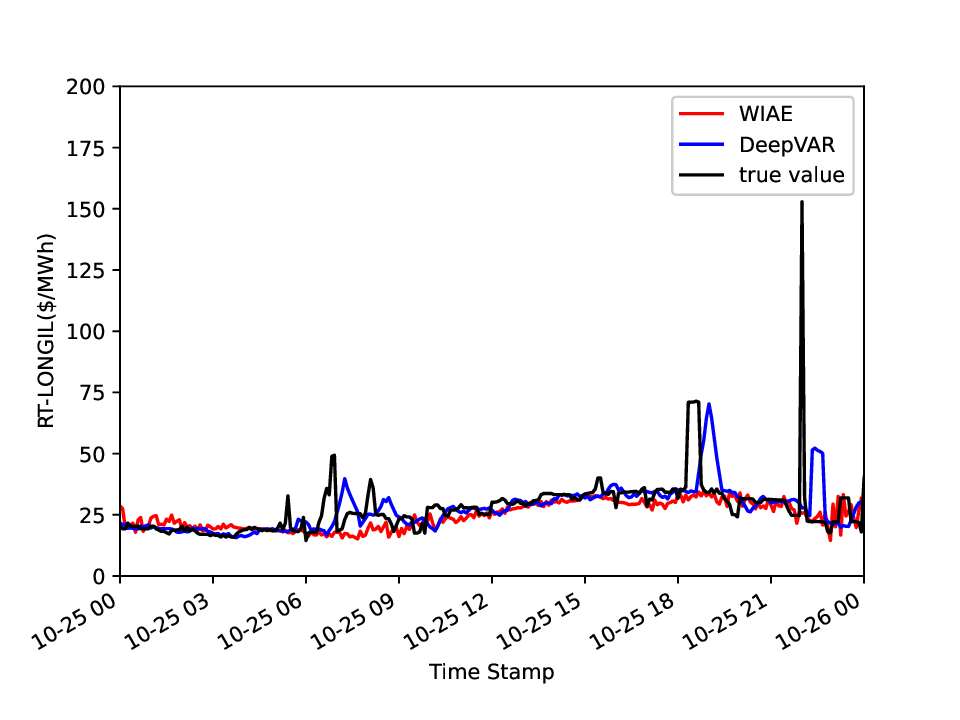}
        \caption{DeepVAR}
        \label{fig:DeepVAR_Fall_Traj}
    \end{subfigure}
    \caption{Trajectories of the real-time price at LONGIL in July (top row) and October (bottom row) 2023, and its prediction generated by selective methods.}
    \label{fig:RTDA_traj}
\end{figure}
\end{landscape}

\begin{landscape}
    \begin{figure}
    \centering
    \begin{subfigure}[t]{0.24\linewidth}
        \includegraphics[width=\linewidth]{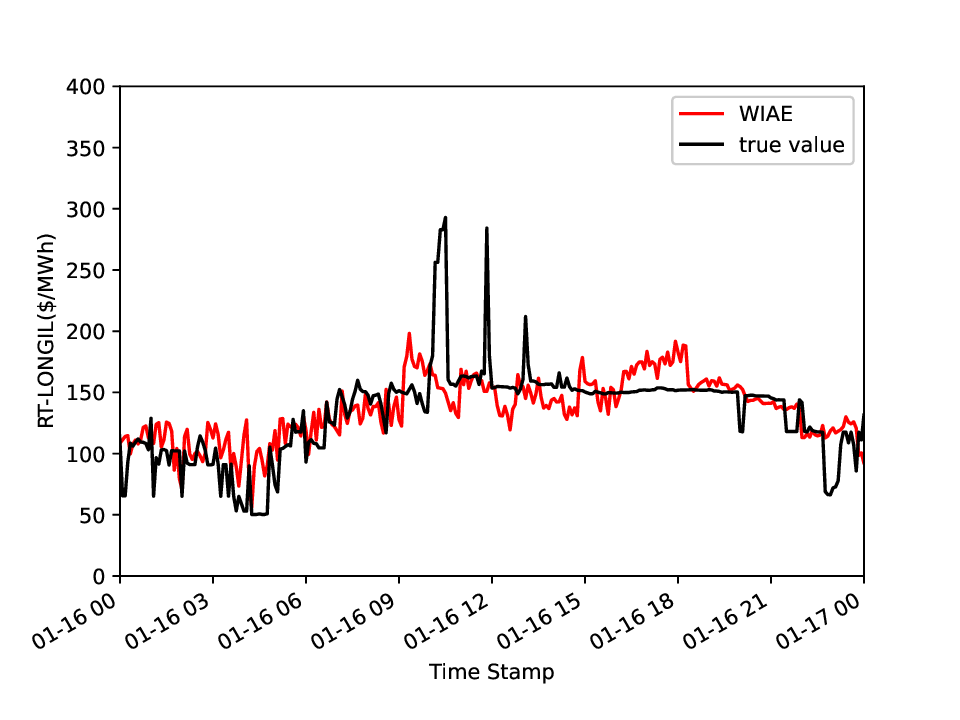}
        \caption{WIAE-GPF}
    \end{subfigure}
    \begin{subfigure}[t]{0.24\linewidth}
        \includegraphics[width=\linewidth]{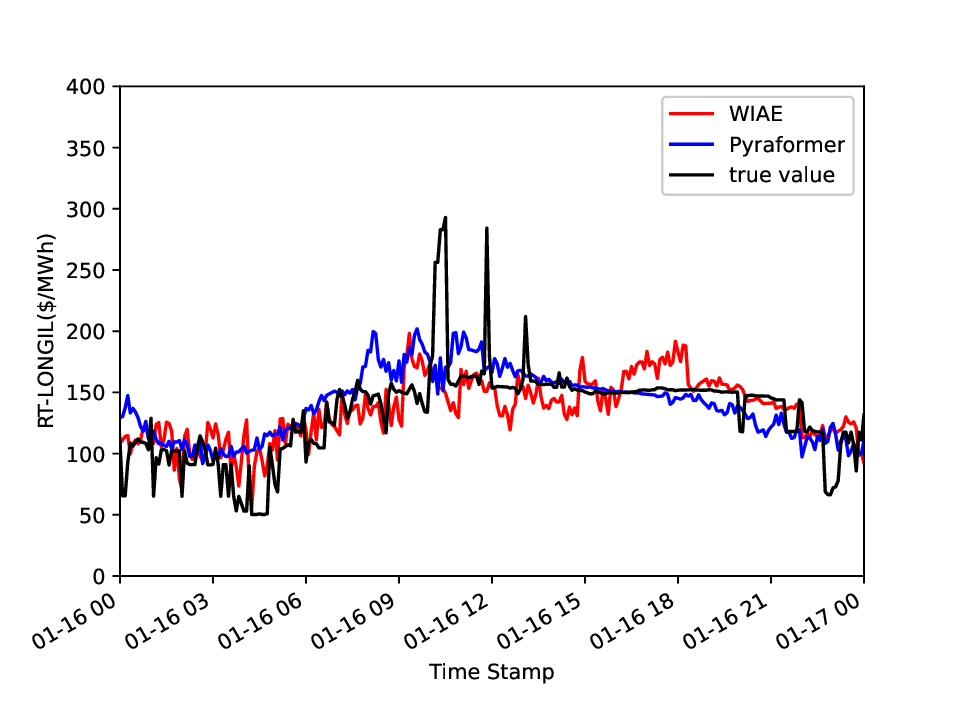}
        \caption{Pyraformer}
    \end{subfigure}
    \begin{subfigure}[t]{0.24\linewidth}
        \includegraphics[width=\linewidth]{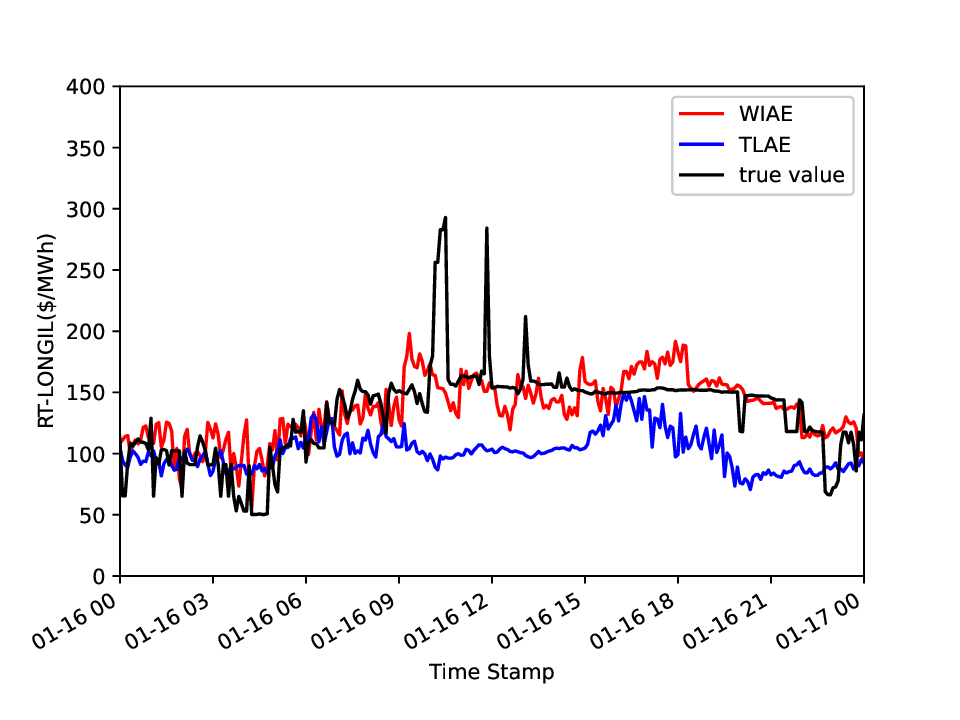}
        \caption{TLAE}
    \end{subfigure}
    \begin{subfigure}[t]{0.24\linewidth}
        \includegraphics[width=\linewidth]{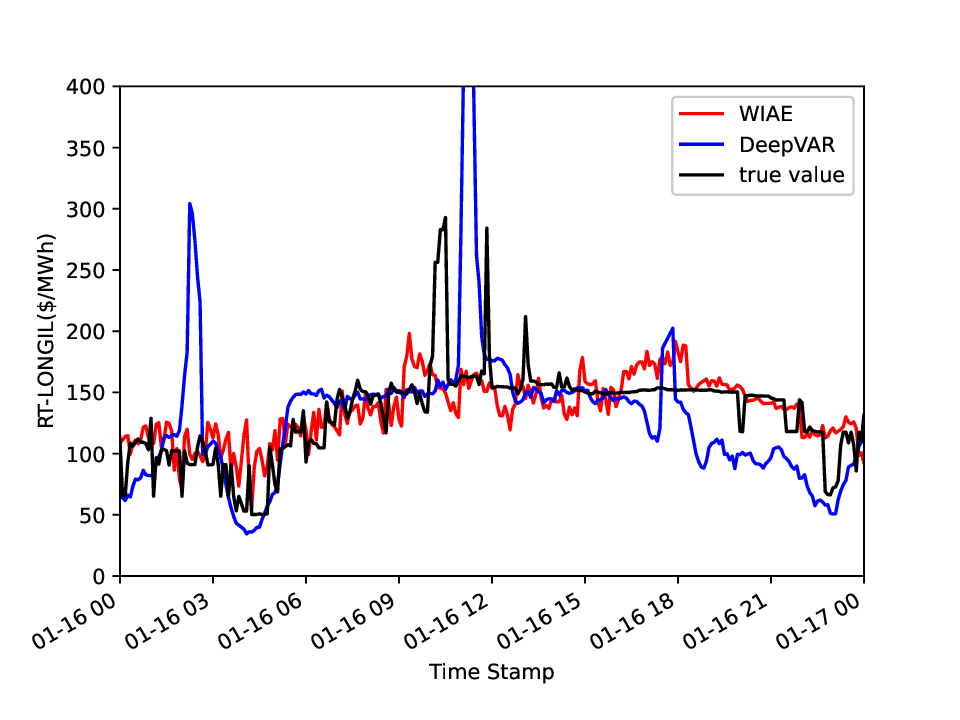}
        \caption{DeepVAR}
    \end{subfigure}
    \\
    \begin{subfigure}[t]{0.24\linewidth}
        \includegraphics[width=\linewidth]{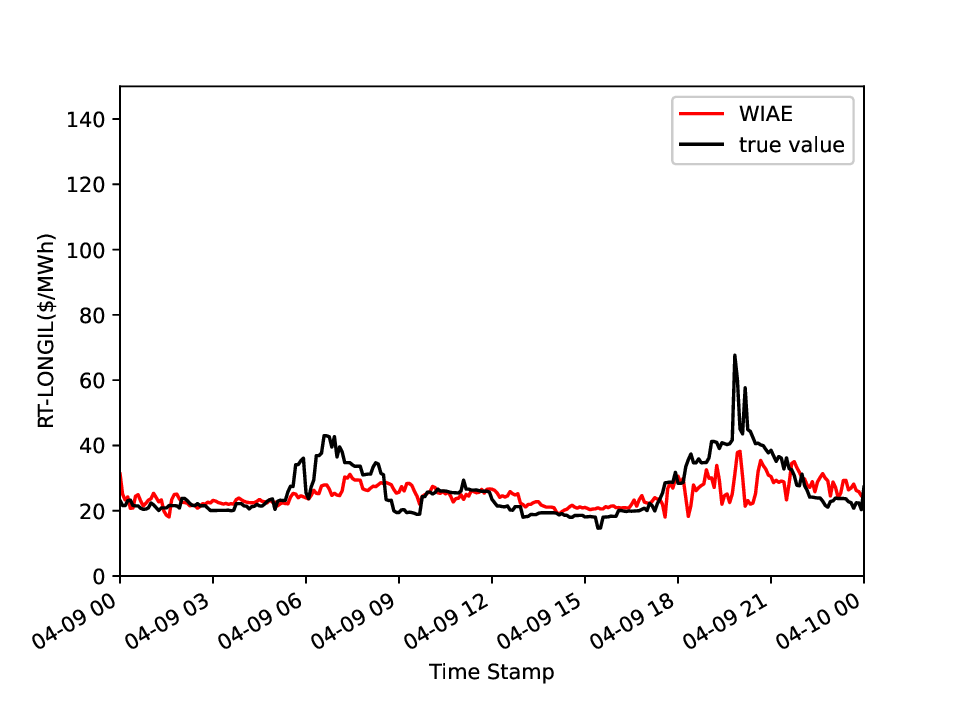}
        \caption{WIAE-GPF}
    \end{subfigure}
    \begin{subfigure}[t]{0.24\linewidth}
        \includegraphics[width=\linewidth]{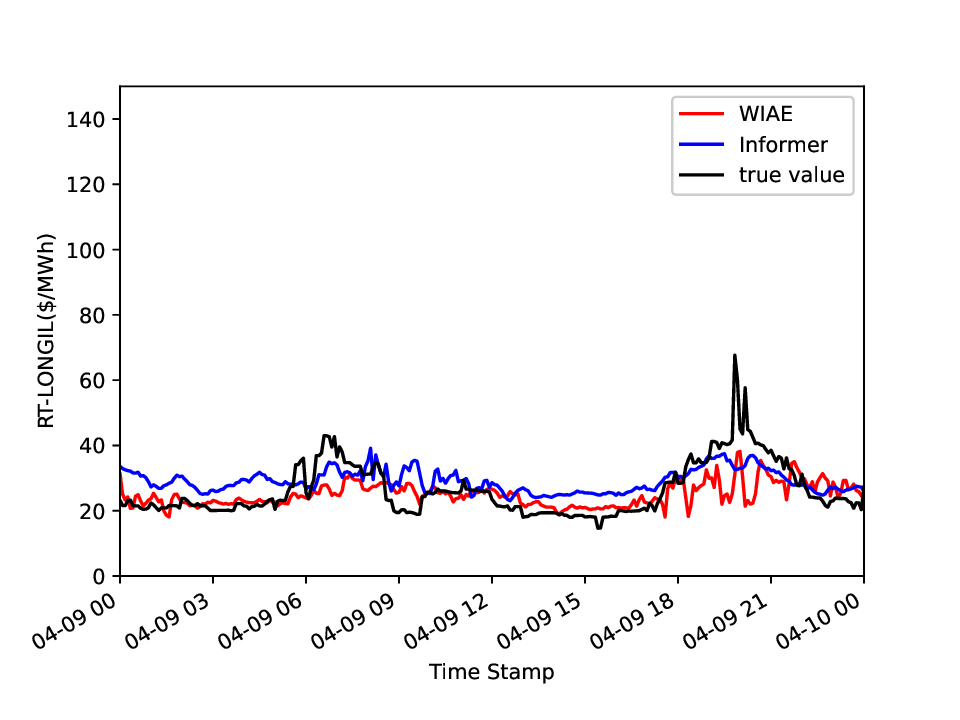}
        \caption{Informer}
    \end{subfigure}
    \begin{subfigure}[t]{0.24\linewidth}
        \includegraphics[width=\linewidth]{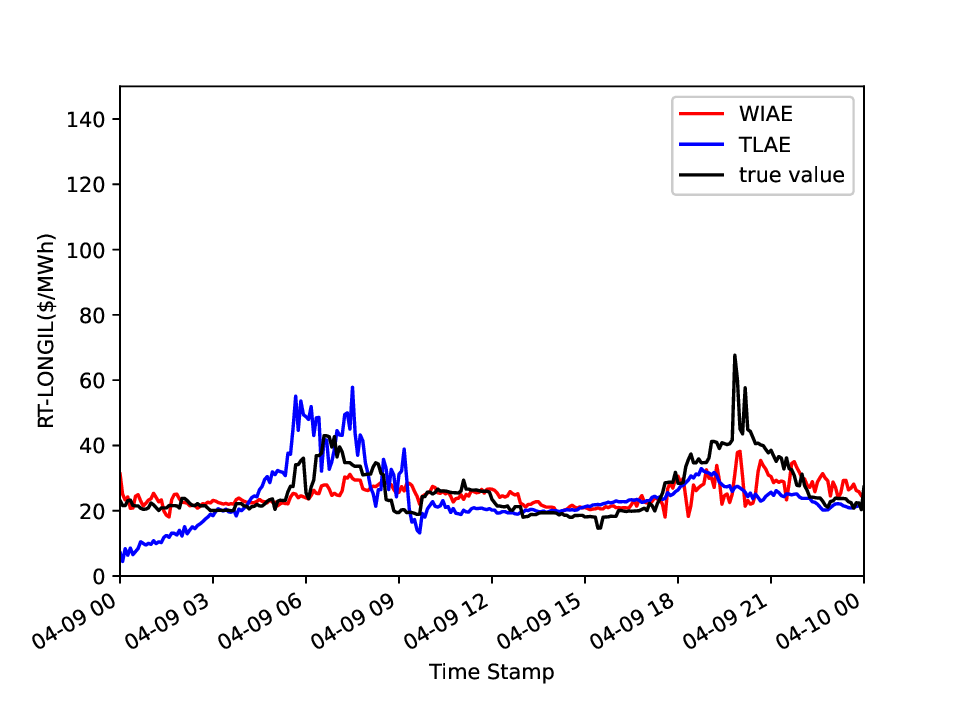}
        \caption{TLAE}
    \end{subfigure}
    \begin{subfigure}[t]{0.24\linewidth}
        \includegraphics[width=\linewidth]{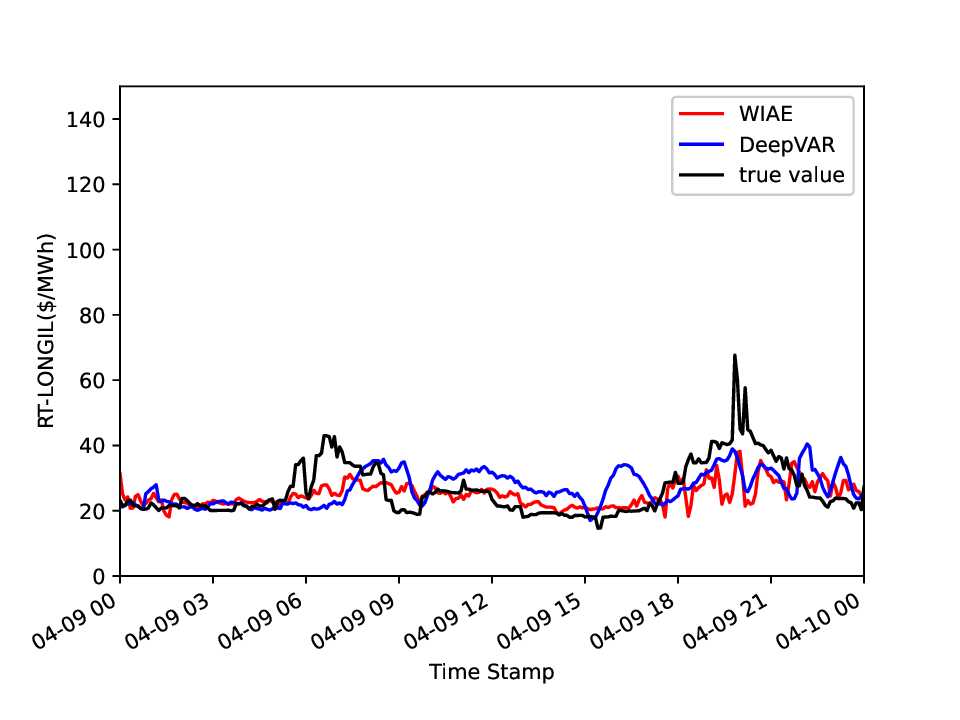}
        \caption{DeepVAR}
    \end{subfigure}
    \caption{Trajectories of the real-time price at LONGIL in January (top row) and April 2024 (bottom row), and its prediction generated by selective methods.}
    \label{fig:RTDA_traj_WS}
\end{figure}
\end{landscape}

To gain insights into the performance of WIAE-GPF and other benchmark techniques, we plotted the ground truth trajectories (black) and trajectory forecasts generated by WIAE-GPF (red) and a competing algorithm (blue) in Fig.~\ref{fig:RTDA_traj} (top row).
Note that the spikes were not predicted by any methods.  
This was not surprising given the nature of how these spikes were produced.  Aside from these spikes, these figures show clearly that WIAE-GPF (red) tracked the ground truth (black) the closest, which was supported by the fact that WIAE-GPF has the smallest NMAE.  We also observed that WIAE-GPF had the smallest variation, which is supported by the fact that WIAE-GPF had the smallest NMSE. 
Furthermore, WIAE-GPF was the least affected by the price spikes. 
This was because, as a GPF method, the Monte-Carlo samples used to produce the MMSE point estimate were less likely to include extreme samples. 

\paragraph{\underline{Fall Real-time LMP Prediction}}

{Test results for October are shown in row 8-13 of Table \ref{tab:RTDA} displayed similarity to summer's results. 
The volatility of fall LMPs is slightly lower than the volatility of summer.
WIAE-GPF performed the best for all except for NMAE, MASE and sMAPE, where it ranked the second.
DeepVAR performed the best for NMAE and MASE, where conditional median is used as the point forecastor.
Pyraformer achieved the best sMAPE and the second-best NMSE.}

{The performance can also be understood by examining the trajectories shown in Fig.~\ref{fig:RTDA_traj} (bottom row). Similarly, WIAE-GPF exhibited the smallest variation and remained closest to the ground truth (excluding the spikes).
In contrast, Fig.~\ref{fig:DeepVAR_Fall_Traj} reveals that DeepVAR tended to predict shifted peaks. This behavior is characteristic of the AR process, which is heavily influenced by past observations.}

\paragraph{\underline{Winter Real-time LMP Prediction}}
{January 2024 has the most volatile LMP among the four months tested. Seen from Table.~\ref{tab:RTDA}, WIAE-GPF obtained the best sMAPE, CRPS, and CPE(90\%), and came close in the second place for the rest. Pyraformer, with its MSE-minimizing training objective, achieved the best NMSE at around $7\%$. DeepVAR had the best NMAE, MASE and CPE(50\%), proving its capability of accurately estimating conditional distribution around the median.}

{Fig.~\ref{fig:RTDA_traj_WS} (top row) showed the trajectory of winter LMP predictions. DeepVAR exhibited larger variability when predicting LMP of January 2024, which explains its higher level of NMSE. Pyraformer and WIAE-GPF achieved similar level of variance when predicting winter LMP. TLAE had the worst ground truth-tracking performance among the four trajectories.}

\paragraph{\underline{Spring Real-time LMP Prediction}}

{April 2024 has the least volatile LMPs among the four months. For this rolling-window experiment, WIAE-GPF achieved the best metrics except for NMAE, MASE and sMAPE. Informer performed better than Pyraformer on the spring dataset, with the best NMAE and MASE. DeepVAR obtained the best sMAPE.}

{Fig.~\ref{fig:RTDA_traj_WS} (bottom row) showed the trajectory of spring LMP predictions. DeepVAR and TLAE exhibited larger deviation for the spring , which explains its higher level of NMSE. Informer exhibited the best ground-truth tracking capability, with WIAE-GPF came in close second.}

\subsection{Interregional LMP spread for Interchange Markets}
\label{subsec:spread}
The interchange market aims to improve overall economic efficiency across ISOs by allowing virtual bidders to arbitrage price differences at proxy buses of two neighboring ISOs.  This experiment was based on the use case of a virtual bidder bidding into the CTS market between NYISO and PJM. The proxy buses of this market were Sandy Point of NYISO and Neptune of PJM.

The CTS market closes 75 minutes ahead of delivery and is cleared every 15 minutes. A virtual bidder submits a price-quantity bid along with the direction of the virtual trade from the source of the proxy with low LMP to the destination proxy with high LMP.  Once the market is cleared, the settlement is based on the actual LMP spread between the two proxies and the cleared quantity.   
The bidder profits if the virtual trade direction matches the direction of the real-time LMP spread. 
Otherwise, the bidder incurs a loss. 
Therefore, the ability to predict the LMP spread direction is especially important.

We performed a 75-minute ahead LMP spread forecasting using the interface power flow and LMP spread data between NYISO and PJM at the Neptune proxy, collected in February 2024.
The interface power flow samples were collected every 5 minutes, and LMP spread every 15 minutes.
We used the first 24 days for training and validation, and the last 5 days of February for testing.

We added {\em Prediction Error Rate (PER)} as a measure for the accuracy of the virtual trading direction prediction, given that the sign of spread is of great importance to profitability. 
PER indicates the percentage of forecasts that don't have the same direction as the ground truth.
For point forecasts, we compared the signs of the forecasts with the signs of the ground truth. 
For probabilistic forecasting, we compared the direction of the ground truth with that of the minimum error-probability prediction of the LMP spread, which is the sign of the conditional median.
For GPF, we compare the sample median with the sign of the ground truth.
\begin{landscape}
    \begin{table}[htbp]
    \centering
    \caption{{\small Evaluation of forecasting results for spread forecasting between NYISO and PJM.}}
    \vspace{0.5em}
    \resizebox{\linewidth}{!}{
    \begin{tabular}{|c|c|c|c|c|c|c|c|c|c|}
    \hline
        {Methods} & {NMSE}  & {NMAE}  &{MASE}  &{sMAPE} &PER  & CRPS & {CPE (90\%)} [NCW] & {CPE (50\%)} [NCW] &CPE (10\%) [NCW]
        \\
        \hline
         WIAE-GPF  &(1) $\mathbf{0.0098}$      &(1) $\mathbf{0.2738}$     &(1) $\mathbf{0.2418}$   &(1) $\mathbf{0.4493}$  &(1) $\mathbf{0.0606}$  &(1) $\mathbf{4.0329}$   
         &(1) $\mathbf{0.0215}$ $[0.4427]$   
         &(2) ${-0.0274}$ $[0.5255]$ 
         &(3) $0.0212$ $[0.1692]$ 
         \\
         \hline
         TLAE \cite{nguyen_temporal_2021}  &(5) $0.9592$   &(5) $0.9785$   &(5) $0.8641$   &(4) $0.4785$ &(4) $0.3692$  &(2) $15.5195$  
         &(3) $-0.0443$ $[0.7745]$  
         &(1) $\mathbf{0.0052}$ $[0.8784]$ 
         &(5) $-0.0933$ $[0.3133]$ 
         \\
         \hline
         DeepVAR \cite{SalinasEtal:19NeuripsDeepVAR}  &(6) {$1.8986$}   &(3) $0.7224$   &(3) {$0.6380$}   &(5) {$0.4806$} &(3) $0.3505$  &(4) {$32.8296$}  
        &(2) $-0.0355$ $[2.0279]$  
        &(3) ${-0.1480}$ $[1.4739]$ 
        &(1) $\mathbf{-0.0021}$ $[0.4198]$
         \\
         \hline
         BWGVT \cite{BottieauEtal:23TPS}  &(3) {$0.9053$} &(4) {$0.8525$}  &(4) {$0.7529$}  &(3) {$0.4674$} &(2) $0.2313$   &(3) {$31.5660$}   
         &(4) {$0.0989$} $[5.1788]$   
         &(4) $0.1835$ $[6.0939]$ 
         &(4) $0.0656$ $[4.0473]$
         \\
         \hline
         Pyraformer \cite{liu_pyraformer_2022} 	&(4) $0.9478$ 		&(6) {$1.2674$} 	&(6) {$1.1193$}  	&(6) {$0.4909$} &(6) $0.6738$  &N/A &N/A  &N/A &N/A
         \\
         \hline
         Informer \cite{ZhouEtal:21AAAI}      &(2) $0.8045$      &(2) $0.4185$       &(2) $0.4252$        &(2) $0.4580$ &(5) $0.5487$  &N/A &N/A  &N/A &N/A
         \\
         \hline
    \end{tabular}}
    \label{tab:spread}

    \caption{{\small Estimation Results of ACE forecasting for PJM. The prediction step is 5-minute ahead.}}
    \vspace{0.5em}
    \resizebox{\linewidth}{!}{
    \begin{tabular}{|c|c|c|c|c|c|c|c|c|}
    \hline
         Methods    &NMSE   &NMAE   &MASE   &sMAPE &CRPS &CPE (90\%) [NCW] &CPE (50\%) [NCW] &CPE (10\%) [NCW]
         \\
         \hline
         WIAE-GPF     &(1) $\mathbf{0.5957}$ &(1) $\mathbf{0.7555}$ &(1) $\mathbf{0.4698}$ &(1) $\mathbf{0.1059}$ &(1) $\mathbf{0.0081}$   
         &(1) $\mathbf{-0.0016}$ $[0.9199]$   
         &(1) $\mathbf{0.0321}$ $[0.9336]$ 
         &(1) $\mathbf{-0.0132}$ $[0.8885]$
         \\
         \hline
         TLAE \cite{nguyen_temporal_2021}      &(5) $1.1727$ &(5) $1.0605$ &(5) $0.6595$ &(3) $0.2782$ &(4) $1.5541$  
         &(4) $-0.7857$ $[0.0004]$  
         &(4) $-0.4489$ $[0.0005]$ 
         &(4) $-0.0957$ $[0.0027]$
         \\
         \hline
         DeepVAR \cite{SalinasEtal:19NeuripsDeepVAR} &(7) $1.4431$ &(7) $1.1750$ &(7) $0.7307$ &(5) $0.3952$ &(3) $1.2947$  
         &(3) $-0.3526$ $[0.5665]$  
         &(3) $-0.2560$ $[0.5296]$ 
         &(3) $-0.0521$ $[0.5434]$
         \\
         \hline
         BWGVT \cite{BottieauEtal:23TPS} &(3) $0.9562$ &(2) $0.9793$ &(2) $0.6090$ &(4) $0.3168$ &(2) $1.2488$  
          &(2) $0.0065$ $[1.8309]$ 
          &(2) $0.0754$ $[2.0385]$ 
          &(5) $0.0996$ $[2.4261]$
         \\
         \hline

         Pyraformer \cite{liu_pyraformer_2022} &(4) $0.9783$ &(4) $0.9948$ &(4) $0.6186$ &(7) $0.4986$ &N/A &N/A &N/A &N/A
         \\
         \hline
         Informer \cite{ZhouEtal:21AAAI} &(2) $0.6006$ &(3) $0.9819$ &(3) $0.6106$ &(2) $0.2247$ &N/A &N/A &N/A &N/A
         \\
         \hline
    \end{tabular}}
    \label{tab:ACE_20}
\end{table}
\end{landscape}

\begin{landscape}
    \begin{figure}
    \centering
    \begin{subfigure}[t]{0.24\linewidth}
        \includegraphics[width=\linewidth]{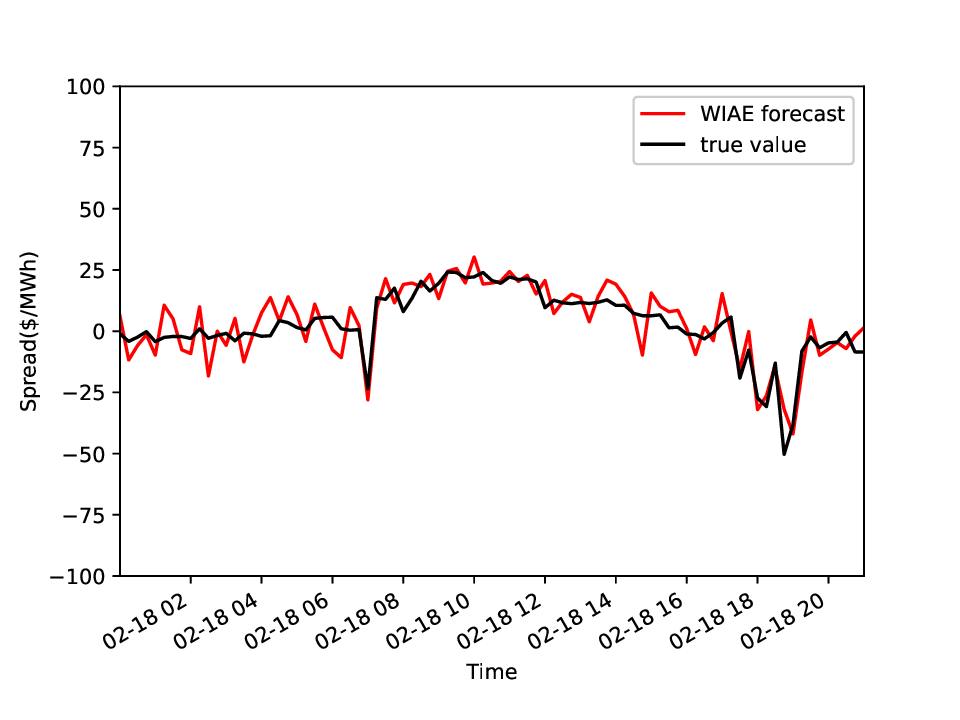}
        \caption{WIAE-GPF}
        \label{fig:WIAE_CTS}
    \end{subfigure}
    \begin{subfigure}[t]{0.24\linewidth}
        \includegraphics[width=\linewidth]{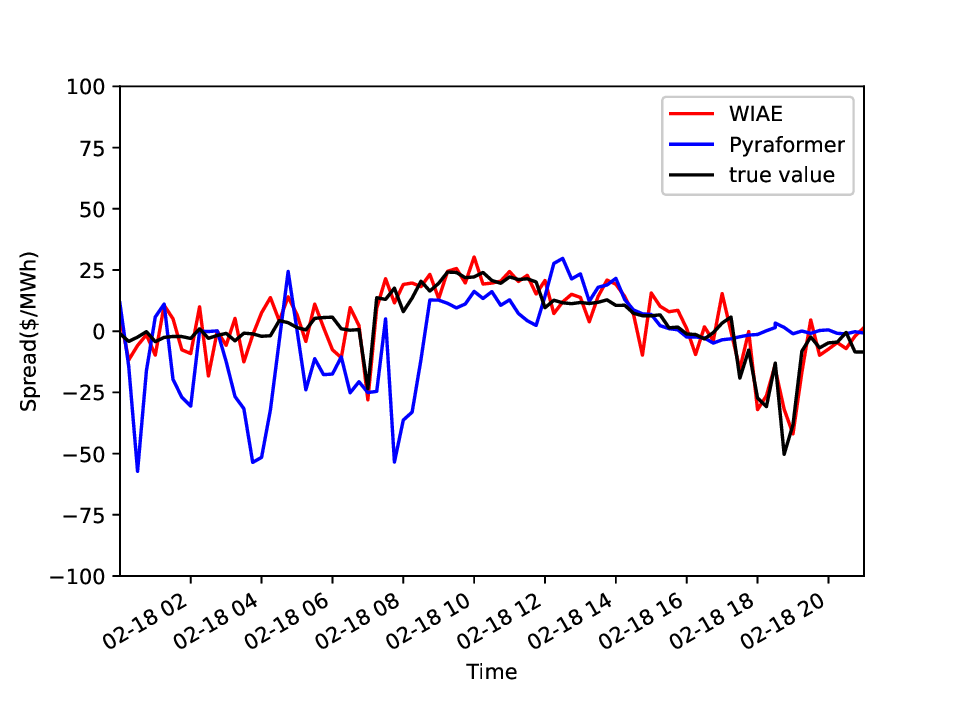}
        \caption{Pyraformer}
        \label{fig:Pyraformer_CTS}
    \end{subfigure}
    \begin{subfigure}[t]{0.24\linewidth}
        \includegraphics[width=\linewidth]{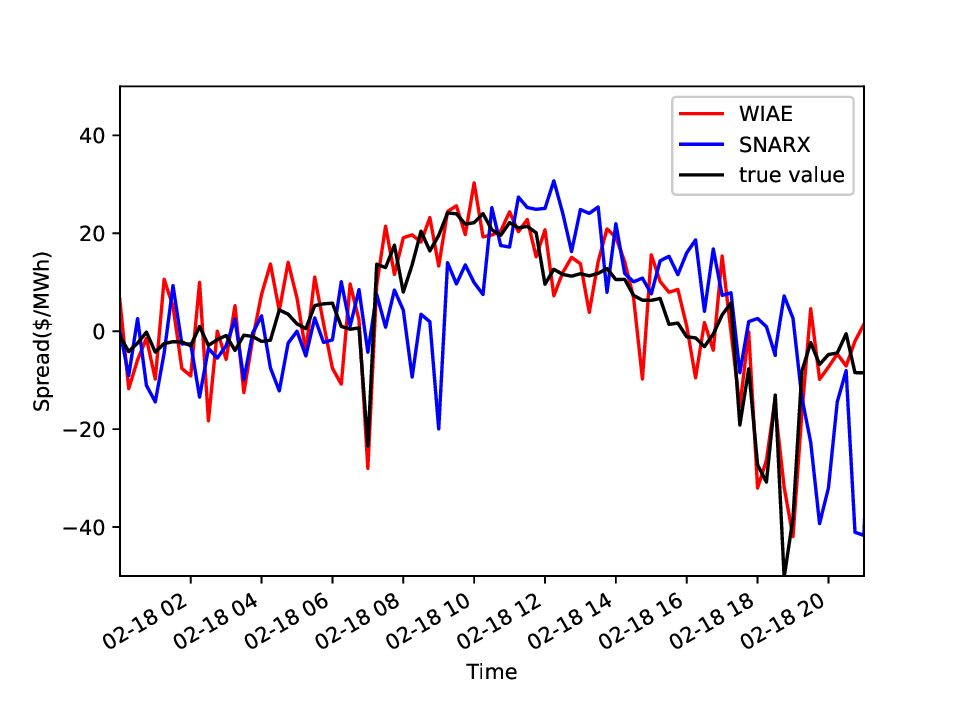}
        \caption{TLAE}
        \label{fig:TLAE_CTS}
    \end{subfigure}
    \begin{subfigure}[t]{0.24\linewidth}
        \includegraphics[width=\linewidth]{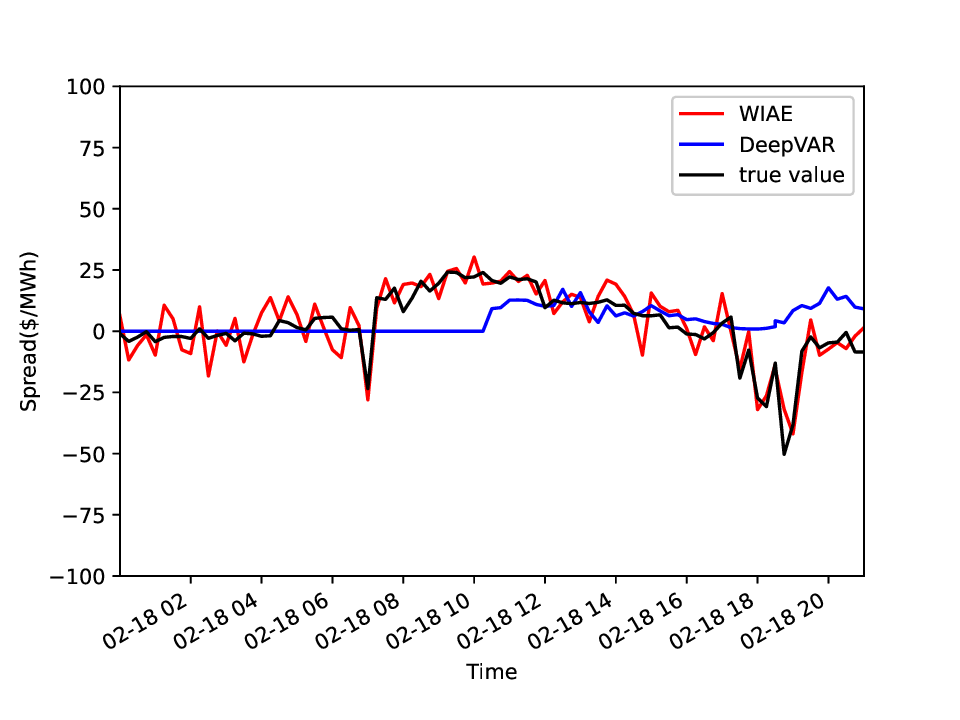}
        \caption{DeepVAR}
        \label{fig:DeepVAR_CTS}
    \end{subfigure}
    \caption{Trajectories of the interregional LMP spread between NYISO and PJM, and its prediction generated by selective methods.}
    \label{fig:CTS_traj}
\end{figure}
\end{landscape}

Seen from Table \ref{tab:spread}, WIAE-GPF performed better than all other techniques in all metrics.
TLAE performed the second-best in CRPS ($15.5195$) but slightly worse than BWGVT when evaluated under point estimation metrics.
Its sequential sampling of the latent Gaussian process added to its numerical instability.
BWGVT was the overall second-best performing probabilistic technique.
Its transformer architecture with enhanced capability of capturing long-term temporal dependency didn't offer much gain for the training difficulty imposed by the increasing number of deep-learning parameters, see Sec.~\ref{subsec:discussion}.
BWGVT also exhibited the tendency to predict a wide interval covering more than the nominal percentage.
Point estimation techniques, namely Pyraformer and Informer, were not competitive when evaluated under point estimation metrics other than NMSE.
Among probabilistic methods, DeepVAR performed similarly to the LLM methods.
The (semi) parametric methods suffered from model mismatch, and were sensitive to sudden changes,
Thus, shifted peaks and valleys were often witnessed in their predictions.

Same observation can also be made through Fig.~\ref{fig:CTS_traj}.
WIAE-GPF has the most stable prediction of interregional LMP spreads, which is corroborated by its smallest NMSE and NMAE.
Pyraformer also follows the trend of LMP spreads accurately but with higher variance.
The AR-based parametric models, DeepVAR exhibited the tendency to predict shifted spikes and failures to catch the rapid and dramatic change of LMP spread.
\subsection{Area Control Error Forecasting for Reserve Market Participants}
ACE is defined as the difference between actual and scheduled load-generation imbalance, adjusted by the area frequency deviation \cite{NERC:11}.  It is the control signal for frequency regulation, and its probabilistic forecasting is especially important for the operator to procure resources and market participants to bid in the regulation ancillary service market.

\begin{figure}[htbp]
    \centering
    \vspace{-1em}
    \includegraphics[width=0.5\linewidth]{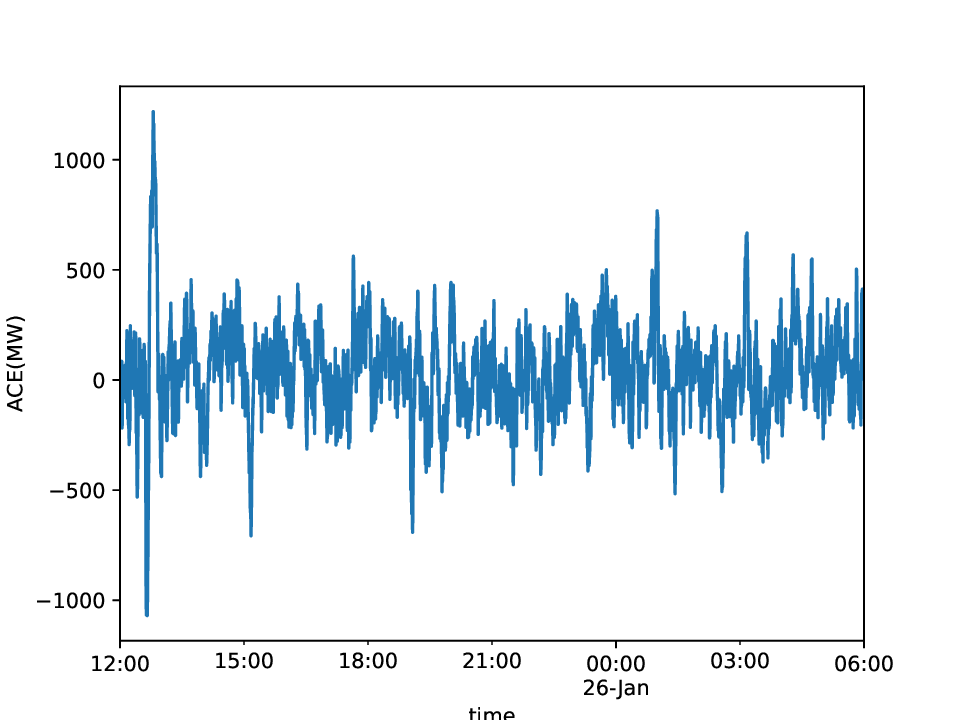}
    \caption{Trajectory of ACE at PJM, Jan. 24th - 26th, 2023.}
    \label{fig:ACE}
\end{figure}

In this subsection, we present the simulation results of a 5-minute ahead forecasting of ACE.
We utilized the ACE data from Jan 24th to 26th, collected by PJM.
The ACE signal is measured every 15 seconds and can be quite volatile, as shown by the trajectory in Fig.~\ref{fig:ACE}.

Shown by Table.~\ref{tab:ACE_20}, WIAE-GPF achieved better performance than other methods, with CRPS less than $0.01$ and sMAPE less than $11\%$, as shown by the WIAE-GPF row.
We credited the strong performance of WIAE-GPF to the simplicity of its latent process, and its Bayesian sufficiency.
BWGVT ranked second among all methods since the ACE data has few outliers.
But its CRPS at $1.2488$ is dramatically larger than that of WIAE-GPF.
Its CPE and NCW for 10\% confidence interval prediction also showed that it cannot accurately predict a narrow interval.
Pyraformer and Informer, trained with NMSE, had better performance under NMSE but worse under NMAE.
With NMSE over $110\%$, TLAE had the worst performance among GPF methods.
DeepVAR and SNARX performed worse than the other forecasting methods, with NMSE and NMAE larger than $110\%$, possibly due to model mismatch.

\vspace{-1em}
\subsection{Discussion: On using LLM for price forecasting}
\label{subsec:discussion}
\begin{table}[htbp]
    \centering
    \caption{\small{Statistics that models long-range dependency of time series.}}
    \vspace{0.5em}
    \begin{tabular}{|c|c|c|c|}
    \hline
         Metrics &Real Time &Interchange Spread & ACE\\
    \hline
        Hurst Exponent  &$0.5257$ &$0.5301$ &$0.5351$\\
    \hline
        DFA &$0.6053$ &$0.6614$ &$0.8609$\\
    \hline
    \end{tabular}
    \label{tab:LR}
\end{table}
The success of LLM-based prediction in natural processing ignited broad interest in adopting LLM models in various applications, including electricity price forecasting with BWGVT, Pyraformer, and Informer. Our experiments showed that the innovation-based method (WIAE) performed uniformly better than the three LLM techniques, except for the real-time prediction at LONGIL-NYC under NMSE, where Pyraformer was the best among all forecasters. Note that the innovation representation used in WIAE can model but not explicitly long-range dependencies of the random process. WIAE does not include attention modeling.

As LLM-based forecasting techniques, Pyraformer, Informer and BWGVT somtimes performed better than TLAE (see {Table~\ref{tab:RTDA}). Compared with the more straightforward deep learning method of DeepVAR, LLM-based forecasting did not show clear advantages.
Authors of \cite{GaneshBunn24TEMPR} pointed out that the simple convolutional neural network outperformed RNNs and LLMs on imbalance price forecasting, for the forecasted time series are not a good fit to the complicated deep learning models.

To understand if long-range dependencies matter in the probabilistic forecasting of electricity market signals, we examined the characteristics of LMP signals using the Hurst exponent and Detrended Fluctuation Analysis (DFA) as indicators for the long-range dependencies of LMP; both parameters had the range [0,1], with deviation from 0.5 indicating symptoms of long-range dependencies.

Table \ref{tab:LR} shows the estimated Hurst exponent and DFA.   The Hurst Exponent and DFA slope displayed a slight deviation from 0.5. The English and Korean literature \cite{MontemurroPury:02Fractals,BhanEtal:06CSF} are known to have long-range dependencies with the Hurst exponents ranging from 0.64 to 0.73. In comparison, the long-term effect of real-time electricity market signals is minimal.

While further studies are necessary, the use of LLM may not be suitable for electricity market signals where long-range dependencies are not evident. 
Indeed, real-time LMPs are computed either on an interval-by-interval basis or as part of short sliding window economic dispatch. Any temporal coupling is a result of temporal dependencies of demand and supplies, neither shown to have long-range dependencies. An unproven hypothesis is that the model complexity of LLM may offset any benefit it may bring to price forecasting.

\section{Conclusion}
\label{sec:conclusion}
This paper introduces WIAE-GPF, a generative AI method for probabilistic forecasting of nonparametric time series, building on the innovation representation developed by Wiener, Kallianpur, and Rosenblatt over six decades ago. Three key findings emerge from our research. First, the innovation representation enables WIAE-GPF to produce accurate conditional probability distributions, provided perfect learning conditions are met. To the best of our knowledge, WIAE-GPF is the first nonparametric generative probabilistic forecasting (GPF) technique with such theoretical guarantees.

Second, WIAE-GPF outperformed leading machine learning-based probabilistic forecasting methods in our numerical experiments using real-world market data. This includes advanced models employing transformer architectures, attention mechanisms, and large language models. Additionally, the local stationarity hypothesis, integral to the weak innovation representation, held robustly in our studies.

Third, this paper establishes Bayesian sufficiency for Rosenblatt’s weak innovation representation, validating it as a canonical framework for time series and a powerful tool for stochastic decision-making. Its potential applications in power system anomaly detection are explored in \cite{Wang&Tong:21JMLR, Kursat&Wang&Tong:21TPS, Tong&Wang&Zhao:24ESJ}.

Finally, deep learning methods are often criticized for their black-box nature, with some viewing them as mysterious techniques that produce impressive yet opaque results. In contrast, WIAE-GPF offers a highly intuitive and interpretable architecture, closely paralleling the classic Kalman filter. Specifically, in Kalman filtering, the innovation is extracted during the measurement update, followed by time-updated predictions based on a state-space model. Similarly, WIAE-GPF extracts innovations using its weak innovation encoder and generates time-updated predictions via the weak innovation decoder. In this sense, WIAE-GPF can be seen as a generalization of Kalman filtering to nonparametric and non-Gaussian settings.

It is important to address the limitations and future directions of this work. WIAE-GPF is derived from the innovation representation of stationary processes, so extending it to certain classes of nonstationary processes would be a natural next step. Notably, an innovation representation exists for nonstationary Gaussian processes using time-varying state-space models. Expanding WIAE-GPF to handle nonstationary time series under regime-switching models is also a promising avenue, particularly given the demonstrated effectiveness of regime-switching techniques in price forecasting. See \cite{GaneshBunn24TEMPR,BottieauEtal:23TPS} for relevant applications.
\appendix
\section{Proof of Theorem~\ref{thm:converge}}
\label{subsec:conv_proof}
Let $\left(\bar{\Vbf}_{t}^{(m)}\right)$ and $\left(\bar{\Xbf}_{t}^{(m)}\right)$ denote the latent process and the reconstruction sequence, under weights $\bar{\theta}_m$ and $\bar{\eta}_m$
\begin{align*}
    \bar{\Vbf}_t^{(m)} = G_{\bar{\theta}}^{(m)}(\Xbf_t,\Xbf_{t-1},\cdots,\Xbf_{t-m+1}),\\
    \bar{\Xbf}_t^{(m)} = H_{\bar{\eta}}^{(m)}(\bar{\Vbf}_t,\bar{\Vbf}_{t-1},\cdots,\bar{\Vbf}_{t-m+1}).
\end{align*}
We define the loss of a WIAE pair $(G_\theta,H_\eta)$ achieved under a $m$-dimensional discriminator pairs as
\begin{multline*}
L^{(m)}(\theta,\eta):=\max_{\gamma,\eta}\big(\mbbE[D_{\gamma}^{(m)}\left(\Ubf_{t:t-m+1}\right)] - \mbbE[D_{\gamma}^{(m)}(\hat{\Vbf}_{t:t-n+1})] \\
     + \lambda(\mbbE[D_{\omega}^{(m)}(\Xbf_{t-n+2:t+T})]
      -\mbbE[D_{\omega}^{(m)}((\Xbf_{t-n+2:t},\hat{\Xbf}_{t+1:t+T}))])\big).
\end{multline*}
We first show that
$L^{(m)}(\theta_m^*,\eta_m^*)\rightarrow 0$ as $m\rightarrow \infty$, where $(\theta_m^*,\eta_m^*)$ denotes the optimal weights of $(G_\theta^{(m)},H_\eta^{(m)})$ obtained by minimizing \eqref{eq:loss}.

Following the line of \cite{Hoffmann-Jørgensen:91Book}, we defined the distance between two random processes $(\Xbf_t)$ and $(\Ybf_t)$ by the expected $\ell_\infty$ norm:
\[d\left((\Xbf_t),(\Ybf_t)\right):= \mbbE\left[\sup_t \lvert \Xbf_t - \Ybf_t \rvert\right].\]
The uniform convergence assumed in assumption A2 is also defined on metric spaces with distance measure $d(\cdot,\cdot)$.
Hence, by assumption A2, $G_{\bar{\theta}}^{(m)}\rightarrow G$ uniformly, which implies that, $\forall \epsilon$, there exists a $M_1$ such that $\forall m > M_1$, $d\left((\bar{\Vbf}_t^{(m)}),(\Vbf_t)\right)<\epsilon.$
Thus, for $\forall F: \ell^\infty(T) \to \mbbR$, $F$ bounded and continuous,
\[d\left(F((\Vbf_t)),F((\bar{\Vbf}_t^{(m)}))\right)<\delta(\epsilon).\]
In other words, 
\[\lim_{m\rightarrow\infty}\mbbE[F((\bar{\Vbf}^{(m)}_t))]=\mbbE[F((\Vbf_t))], \]
which fulfills the definition of weak convergence.
Therefore,
\begin{align}
    \bar{ \Vbf}_{t:t-m+1}^{(m)}\stackrel{\mbox{\tiny d}}{\rightarrow}\Vbf_{t:t-m+1},
    \label{eq:G_conv}
\end{align}
due to the fact that convergence in expectation implies convergence in distribution. 

Similarly, by the uniform convergence of $H_{\bar{\eta}}^{(m)}$ to $H$, we have that that $\forall m >M_2$,
\begin{align*}
d\left((\hat{\Xbf}_t),(H_{\bar{\eta}}^{(m)}((\Vbf_t)))\right) <\epsilon,
\end{align*}
where $(H_{\bar{\eta}}^{(m)}((\Vbf_t))))$ represent the random sequence generated by passing $(\Vbf_t)$ through $H_{\bar{\eta}}$.
Thus, for $\forall F: \ell^\infty(T) \to \mbbR$, $F$ bounded and continuous,
\[d\left(F((\hat{\Xbf}_t)),(H_{\bar{\eta}}^{(m)}((\Vbf_t))))\right)<\delta(\epsilon).\]
Hence we have $(H_{\bar{\eta}}^{(m)}((\Vbf_t))))$ converges in distribution to $(\hat{\Xbf}_t)$.
Since $H$ is continuous and $H_{\bar{\eta}}^{(m)}$ converges uniformly to $H$, $H_{\bar{\eta}}^{(m)}$ is also continuous. 
Thus, by continuous mapping theorem, 
\[\bar{ \Vbf}_{t-m+1:t}^{(m)}\stackrel{\mbox{\tiny d}}{\rightarrow}\Vbf_{t-m+1:t}\stackrel{}{\Rightarrow}H_{\bar{\eta}}^{(m)}(\bar{ \Vbf}_{t-m+1:t}^{(m)})\stackrel{\mbox{\tiny d}}{\rightarrow}H_{\bar{\eta}}^{(m)}\left(\Vbf_{t-m+1:t}\right),\]
that is, $\bar{\Xbf}_{t}^{(m)}\stackrel{\mbox{\tiny $d$}}{\rightarrow}H_{\bar{\eta}}^{(m)}(\Vbf_{t},\cdots,\Vbf_{t-m+1})$. 
Therefore, 
\begin{align}
    (\Xbf_{t-m+2:t},\bar{\Xbf}_{t+T}^{(m)})\stackrel{\mbox{\tiny d}}{\rightarrow} (\Xbf_{t-m+2:t},\hat{\Xbf}_{t+T})\stackrel{\mbox{\tiny d}}{=}(\Xbf_{t-m+2:t},\Xbf_{t+T}).
    \label{eq:H_conv}
\end{align}
By \eqref{eq:G_conv}\&\eqref{eq:H_conv}, $L^{(m)}(\bar{\theta}_m,\bar{\eta}_m)\rightarrow 0$.
Since $\theta^*_m$ and $\eta^*_m$ are the optimal parameters obtained by minimizing \eqref{eq:loss} evaluated by $m$-dimensional discriminators $(D_\omega^{(m)},D_\gamma^{(m)})$,
\begin{align*}
L^{(m)}(\theta_m^*,\eta_m^*):=\min_{\theta,\eta}L^{(m)}(\theta,\eta) \leq L^{(m)}(\bar{\theta}_m,\bar{\eta}_m)\rightarrow 0.
\end{align*}

Because $L^{(m)}(\theta_m^*,\eta_m^*)\rightarrow 0$ as $m\rightarrow \infty$, $\boldsymbol V_{t:t-m+1}^{(m)} \stackrel{\mbox{\tiny d}}{\rightarrow} \boldsymbol V_{t:t-m+1}^{(m)}$ and $(\Xbf_{t-m+2:t},\hat{\boldsymbol X}_{t+T}^{(m)})\stackrel{\mbox{\tiny d}}{\rightarrow}\boldsymbol (\Xbf_{t-m+2:t},\boldsymbol X_{t+T}^{(m)})$ follow directly from the equivalence of convergence in Wasserstein distance and convergence in distribution \cite{Villani09:Book}.
Since the discriminator dimensionality also goes to $\infty$, we have $(\Xbf_{0:t},\hat{\Xbf}_{t+T}^{(m)})\stackrel{\mbox{\tiny d}}{\rightarrow}(\Xbf_{0:t},\Xbf_{t+T})$.
Further, the conditional distribution of $\hat{\Xbf}_{t+T}^{(m)}|\Xbf_{0:t}=\xbf_{0:t}$ converges in distribution to $\Xbf_{t+T}|\Xbf_{0:t}=\xbf_{0:t}$ follows from a simple application of the Bayes rule. $\square$
\section{Definition of Metrics for Time Series Forecasting}
\label{sec:metrics}
Given the original time series $(\xbf_t)$, the forecasts $(\tilde{\xbf}_t)$, $N$ the size of datasets, and $T$ the prediction step, the point estimation metrics can be calculated through:
\begin{align*}
    \mbox{NMSE} = \frac{\frac{1}{N-T}\sum_{t=T+1}^N(\xbf_{t}-\tilde{\xbf}_{t-T})^2}{\frac{1}{N-T}\sum_{t=T+1}^N \xbf_t^2},\\
    \mbox{NMAE} = \frac{\frac{1}{N-T}\sum_{t=T+1}^N|\xbf_t-\tilde{\xbf}_{t-T}|}{\frac{1}{N-T}\sum_{t=1}^N|\xbf_t|},\\
    \mbox{MASE} = \frac{\frac{1}{N-T}\sum_{t=T+1}^N|\xbf_t-\tilde{\xbf}_{t-T}|}{\frac{1}{N-T}\sum_{t=T+1}^N|\xbf_t-\xbf_{t-T}|},\\
    \mbox{sMAPE} = \frac{1}{N-T}\sum_{t=1}^N\frac{|\xbf_t-\tilde{\xbf}_{t-T}|}{(|\xbf_t|+|\tilde{\xbf}_{t-T}|)/2}.
\end{align*}
The purpose of adopting multiple metrics is to comprehensively evaluate the forecasting performance.
NMSE and NMAE evaluate the overall performance, and MASE reflects the relative performance to the naive forecaster.
Methods with MASE smaller than $1$ outperform the naive forecaster.
sMAPE is the symmetric counterpart of mean absolute percentage error (MAPE) that can be both upper bounded and lower bounded.
Since for electricity datasets, the actual values can be very close to $0$, thus nullifies the effectiveness of MAPE, we regard sMAPE as the better metric.

For probabilistic methods, we further evaluates their CRPS.
CRPS can be computed from
\begin{align*}
    \mbox{CRPS}= \frac{1}{N-T}\sum_{t=T+1}^{N}\left(\int_\mathbb{R}\left( \tilde{F}(\xbf|\xbf_{1:t-T}) -\mathbb{I}_{\xbf_{t}\leq\xbf}\right)^2d\xbf\right),
\end{align*}
where $\mathbb{I}$ is the indicator function and $\tilde{F}(\xbf|\xbf_{0:t-T})$ the empirical cumulative density function (c.d.f.) of $\tilde{\Xbf}_{t-T}$ conditioned on $\Xbf_{0:t-T}=\xbf_{0:t-T}$ predicted by probabilistic forecasting methods.
CRPS is equivalent to comparing the empirical conditional c.d.f. forecasted by probabilistic methods with the indicator c.d.f.  $\mathbb{I}_{\tilde{\xbf}_{t-T}>\xbf_t}$ of the true value $\xbf_t$.
It can be viewed as a generalization of MAE to probabilistic methods.

The coverage probability (CP) of an confidence interval predictor is the (estimated) probability that the ground truth falls within the predicted interval.
For a $T$-step prediction of $\beta\%$-intervals, we denote the upper and lower bound by $\hat{U}_{t|t-T,\beta}$ and $\hat{L}_{t|t-T,\beta}$.
CP can be computed through
\begin{align*}
    \mbox{CP}(\beta\%) =\frac{1}{N-T} \sum_{t=T+1}^{N}\mathbb{I}_{\xbf_{t}\in[\hat{L}_{t|t-T,\beta},\hat{U}_{t|t-T,\beta}]}.
\end{align*}
The closer the CP to its nominal value $\beta\%$, the more accurate the prediction is.
Thus, the coverage probability error (CPE) is often adopted for evaluation.
CPE measures the deviation of CP from its nominal value $\beta\%$
\[\mbox{CPE}(\beta\%) = \mbox{CP}(\beta\%) - \beta\%.\]
The value of CPE closer to zero means the prediction interval estimation is more accurate.

Although CP and CPE are widely adopted for its simplicity, since they only estimate the unconditional coverage, they do not measure the accuracy the coverage based on the forecasted conditional probability distribution.
Its limitation was discuss in \cite{christoffersen_evaluating_1998}.

In particular, while a good forecaster produces small CPE and a forecaster with high CPE must be a poor forecaster,  a forecaster producing small CPE may not be a good forecaster.  To this end, the normalized coverage width (NCW) can be used as a secondary measure.
NCW is defined as 
\begin{align*}
    \mbox{NCW}(\beta\%) = \frac{1}{N-T}\sum_{t=T+1}^{N}\frac{\hat{U}_{t|t-T,\beta}-\hat{L}_{t|t-T,\beta}}{\hat{U}_\beta - \hat{L}_\beta},
\end{align*}
where $\hat{U}_\beta$ and $\hat{L}_\beta$ are the prediction interval estimated from the empirical quantile of the testing data.
For instance, when predicting a $90\%$ interval, $\hat{U}_{90}$ is the empirical $0.95$-quantile of the testing set, whereas $\hat{L}_{90}$ is the empirical $0.05$-quantile of the testing set. 
As a result, NCW is the average width of intervals predicted normalized by the width of the interval estimated through the empirical marginal distribution of the testing set.
One would expect that, conditional on observations, one would get a more concentrated prediction interval than the interval estimated based on unconditional distribution.
Hence, a method with NCW smaller than $1$ estimates prediction interval more accurately than the unconditional estimation.
At similar level of CP, the method with smaller NCW shows better accuracy in prediction interval estimation.

 \bibliographystyle{elsarticle-num} 
 \bibliography{EPF}






\end{document}